\documentclass{aa}  
\usepackage{graphicx}
\usepackage{natbib}
\bibpunct{(}{)}{;}{a}{}{,} % to follow the A&A style
\usepackage{txfonts}
\begin{document}

   %\title{Chemical Evolution of the Alpha-Rich  Galactic Bulge Globular Cluster NGC 6440}

 \title{The peculiar Na-O anticorrelation of the bulge globular cluster NGC~6440}

   %\subtitle{I. Overviewing the $\kappa$-mechanism}
   \author{C. Mu\~{n}oz\inst{1,2} , S. Villanova\inst{1}, D. Geisler\inst{1}, I. Saviane\inst{2}, B. Dias\inst{2}, R.E. Cohen\inst{1} \& F. Mauro\inst{1}}

  % \author{G. Wuchterl
    %      \inst{1}
     %     \and
      %    C. Ptolemy\inst{2}\fnmsep\thanks{Just to show the usage
       %   of the elements in the author field}
        %  }

   \institute{Departamento de Astronom\'{i}a, Casilla 160-C, Universidad de
  Concepci\'{o}n, Concepci\'{o}n, Chile\\
  \email{cesarmunoz@astro-udec.cl}
  \and European Southern Observatory, Casilla 19001, Santiago, Chile
              }
              
\titlerunning{The Peculiar Na-O anticorrelation  of NGC 6440}
\authorrunning{C. Munoz et al.}

  %%          T\"urkenschanzstrasse 17, A-1180 Vienna\\
   %          \email{wuchterl@amok.ast.univie.ac.at}
   %      \and
    %         University of Alexandria, Department of Geography, ...\\
     %        \email{c.ptolemy@hipparch.uheaven.space}
      %       \thanks{The university of heaven temporarily does not
       %              accept e-mails}
        %     }

   \date{}

% \abstract{}{}{}{}{} 
% 5 {} token are mandatory
 
  \abstract
  % context heading (optional)
  % {} leave it empty if necessary  
   {Galactic Globular Clusters (GCs)  are essential tools to understand  the earliest epoch of    the Milky Way, since they are among the oldest objects in the Universe and can be used to trace its formation and evolution.\\
 Current studies using high resolution spectroscopy for many stars in each of a large sample of GCs allow us  to develop  a detailed observational picture about their formation and their relation with the Galaxy. However, it is  necessary  to complete  this picture by including GCs that belong to all major Galactic components, including the Bulge.}
  % aims heading (mandatory)
   {Our aim is to perform a detailed chemical analyses of the bulge GC NGC 6440 in order  to determine if this object has Multiple Populations (MPs) and  investigate  its relation with the Bulge of the Milky Way and with the other Galactic GCs, especially those associated  with the Bulge,  which are largely poorly studied.}
  % methods heading (mandatory)
   {We determined the stellar parameters and  the chemical abundances of light elements (Na, Al), iron-peak elements (Fe, Sc, Mn, Co, Ni),  $\alpha$-elements (O, Mg, Si, Ca, Ti)  and heavy elements (Ba, Eu) in seven red giant members of NGC 6440  using high resolution spectroscopy from FLAMES@UVES.}
  % results heading (mandatory)
   { We found a mean iron content of [Fe/H]=-0.50$\pm$0.03  dex in agreement with other studies. We found no internal iron spread. On the other hand, Na and Al show a significant intrinsic spread, but the cluster has no significant O-Na anticorrelation nor exhibits a Mg-Al anticorrelation.  The $\alpha$-elements show good agreement with the Bulge field star trend, although they are at the high alpha end and are also higher than those of other GCs of comparable metallicity. The heavy elements are dominated by the r-process, indicating a strong contribution by SNeII. The chemical analysis suggests an origin similar to that of the Bulge field stars.}
  % conclusions heading (optional), leave it empty if necessary 
   {}

\date{Accepted}
   \maketitle
%
% * <fullchileaa@gmail.com> 2016-12-09T03:08:33.822Z:
%
% ^.
% * <fullchileaa@gmail.com> 2016-12-09T03:08:30.892Z:
%
% ^.
%________________________________________________________________
\section{Introduction}

GCs are ideal laboratories to study stellar evolution because  they are among the oldest known objects in the universe. They  have been extensively studied with photometry in a wide variety of different photometric systems, and with high, medium, and low resolution spectroscopy.

They  were until recently thought to be Simple Stellar Populations (SSP), with all of  the stars in a cluster having the same age and initial chemical composition. However, they have  proven to be  far more complex objects. For example, most  Galactic GCs have recently been shown to be characterized by in-homogeneities in the light-element content (C, N, O, Na, Mg  and/or Al),  that translate in well defined patterns such as O-Na or Mg-Al anticorrelations. This spread is most likely due  to the early self-pollution the cluster suffered after its formation, allowing the birth of two  or more generations of stars \citep{gratton04}.
Several kinds of polluters for the light elements have been proposed: intermediate mass AGB stars \citep{dantona02,dantona16}, fast rotating massive MS stars \citep{decressin07,krause13} and massive binaries \citep{demink10,izzard13}.\\
%In addition, several massive GCs 
 In addition, a few massive GC-like systems such as $\omega$ Cen \citep{johnson08,marino11a}, M22 \citep{marino11b,dacosta09}, M54 \citep{carretta10a} and  Terzan 5 \citep{origlia11} show a  significant spread in Fe as well. Such clusters must also have been able to retain material ejected from SNeII and/or SNeIa \citep{marcolini09}, as well as material ejected from polluters at lower velocity  via stellar winds. This material  ejected into the  interstellar medium of the GC  mixes with the primordial gas to form the second generation of stars, if the appropriate conditions arise \citep[e.g.][]{cottrell81, carretta10b}. However, none of the current scenarios are able to satisfactorily reproduce the complex multiple population behavior now known to exist \citep[e.g.][]{renzini15}.

Despite the large number of galactic GCs that have been studied with high resolution spectroscopy, which allows a more detailed understanding of their evolution, the picture is not yet complete, because  many galactic GCs, especially those belonging to the Bulge of the Milky Way, have not been studied in detail yet,  mainly because of the large and often variable reddening, as well as intense field star contamination, which plague detailed studies.

In this paper we present a  detailed chemical study of the metal rich  GC NGC 6440, located 1.3 kpc away from the center of our Galaxy \citep[2010 edition]{harris96}. We have analyzed using high resolution spectroscopy the chemical patterns of seven  member stars and obtained the abundances of light, $\alpha$,  iron-peak   and  heavy elements. The detailed abundance distribution of these elements is compared to that of field stars in several major Galactic components (bulge, disk, halo) as well as other (bulge and non-bulge) GCs.\\
% * <fullchileaa@gmail.com> 2016-11-21T15:34:13.452Z:
%
% ^.
% * <fullchileaa@gmail.com> 2016-11-21T15:34:08.699Z:
%
% ^.

NGC 6440 is particularly interesting because \citet{mauro14} suggested the presence  of a possible iron spread in this cluster, based on low resolution Ca triplet spectra of 8 stars. On the other hand \citet[hereafter OR08]{origlia08} have found a homogeneous iron content in a sample of 10 stars using high resolution infrared spectroscopy. In this paper we will address  this discrepancy.

In addition, \citet{mauro12} found in NGC 6440 two  horizontal branches (HBs) using data from the VVV survey \citep{minniti10}. 

 This peculiar feature has been  found  in  only  three other bulge stellar systems, namely Terzan 5 \citep{ferraro09}, NGC 6440 and NGC 6569 \citep{mauro12}.
While Terzan 5 has been extensively studied in  recent years and turned out to be a complex stellar system with multiple stellar populations with different [Fe/H] \citep{origlia11,massari14} and age \citep{ferraro16}, NGC 6440 and NGC 6569 still need to be fully characterised.

%Several other bulge GCs like NGC 6569 \citep{mauro12} and Terzan 5 \citep{ferraro09} show a similar behavior. This peculiarity has only been found in these three Bulge GCs, which are among the most massive. This is  an indication of the presence of multiple populations \citep{rood85}, which we want to characterize chemically.

Several chemical analyses have been carried out on NGC 6440 stars using different techniques. However, these studies are far from being complete. For example, OR08 measured light and $\alpha$ elements using IR high resolution spectroscopy (see Table \ref{ces-ori} and Figure \ref{comp-OR08}). Moreover, there are other studies with low resolution spectroscopy and photometry \citep{valenti04,dias16a}. Our sample, although slightly  smaller than the previous studies, was observed with high resolution, which allows us to obtain the abundances of fourteen  elements with smaller errors than the previous studies.

In section 2 we describe the observations and data reduction, in section 3 we explain the methodology we used to calculate  atmospheric parameters and chemical abundances.
In section 4 we present our result concerning  iron-peak elements, alpha elements, the Na-O anticorrelation, the Mg-Al and O-Al relations and heavy element abundances.
Our findings are used in Sec. 5 to shed light on the origin of NGC 6440, and finally in Section 6 our conclusions are presented.

%__________________________________________________________________
\section{Observations and data reduction}
 Bright red giant star candidate stars in NGC 6440 were observed with the fiber-fed multi-object FLAMES facility mounted at the ESO VLT/UT2 telescope in Cerro Paranal (Chile) in period 93A (ESO program ID 093.D-0286, PI S. Villanova). The FLAMES observations analysed here were conducted using the blue and red arms of the high resolution spectrograph UVES and allowed the simultaneous observations of  seven stars.

\begin{figure}
\centering
  \includegraphics[width=3.5in,height=3.3in]{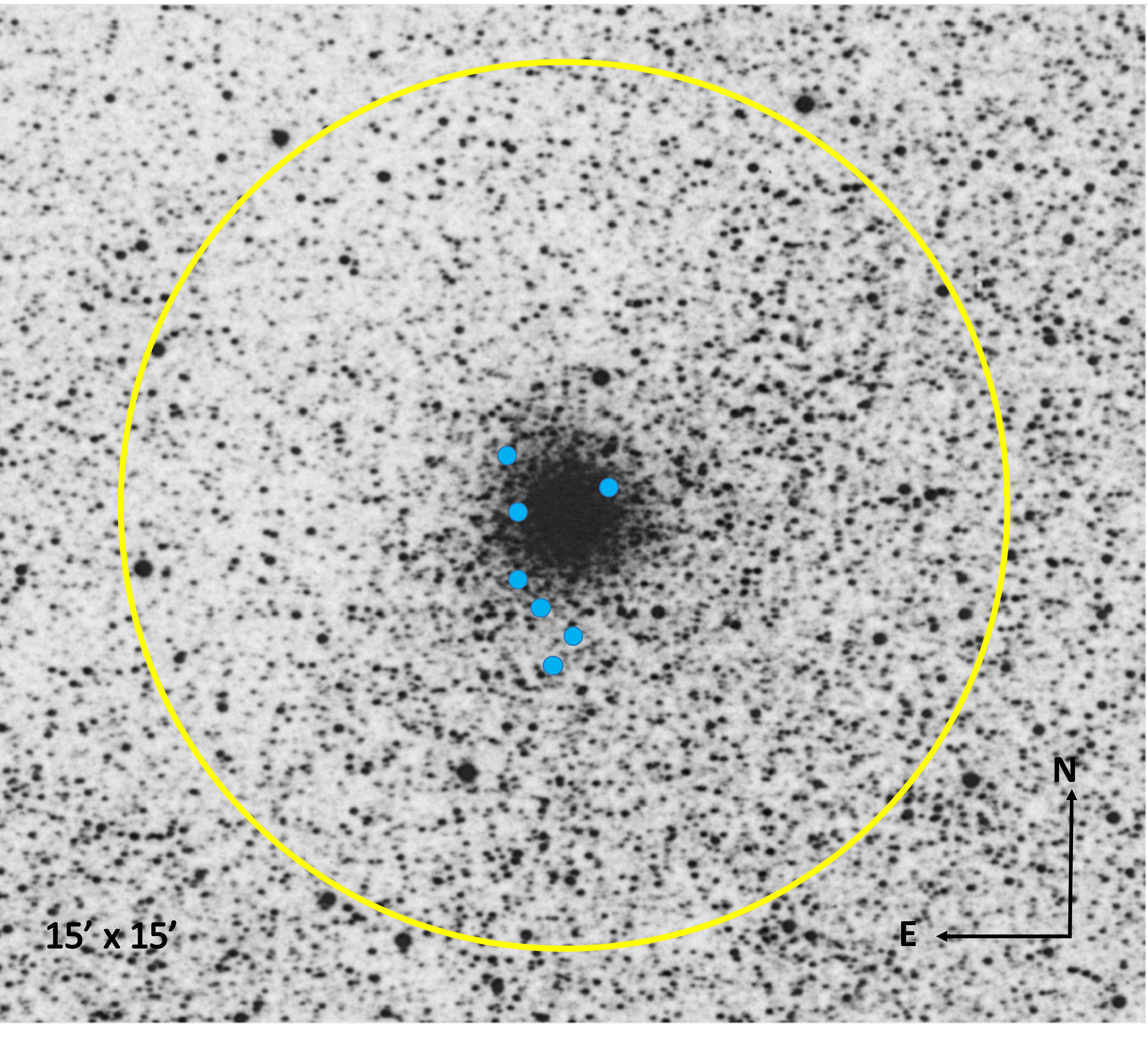}
  \caption{Distribution of  the stars observed  in  NGC 6440 (blue filled circles). The yellow large  circle is the tidal radius \citep[2010 edition]{harris96}.}
  \label{spatial}
 \end{figure}

We selected seven targets to be observed with FLAMES@UVES from the membership list of NGC 6440 previously published in \citet{saviane12} using FORS2: their spatial distribution is shown in Figure \ref{spatial}. \citet{saviane12}  showed that the stars used in this study are members of NGC 6440 using two criteria: The range of radial velocities of member stars was small and the dispersion in the equivalent widths was comparable to the measurement  errors (assuming the intrinsic abundance dispersion in the cluster is small).

\begin{figure}
\centering
  \includegraphics[width=3.5in,height=4.1in,angle=-90]{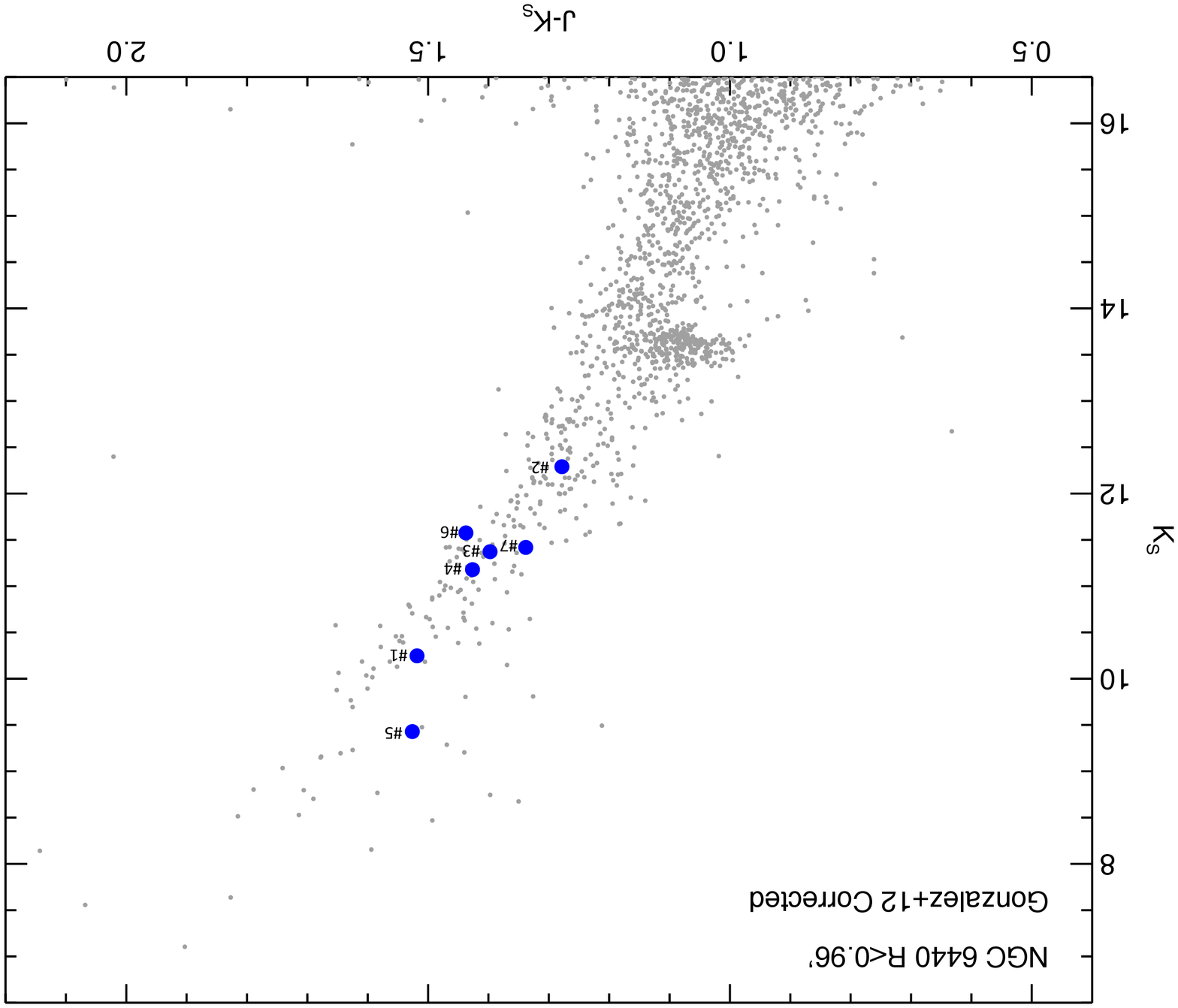}
  \caption{CMD of  NGC 6440 from the VVV survey  corrected  by   the VVV reddening maps \citep{gonzalez12}. The blue filled circles are our observed UVES sample.}
  \label{CMD}
 \end{figure}

  These stars belong to the upper  RGB, as can be clearly seen in the CMD of the cluster (Figure \ref{CMD}), with the exception of  star \#5 , which apparently is an AGB star. Its chemical pattern shows good agreement with the rest of the sample.

%----------------------------------------------------------------
 
%----------------------------------------------------------------
%----------------------------------------------------------------
FLAMES-UVES data have a spectral resolution of about R$\simeq$47000. The data was taken with central wavelength 580\,nm, which covers the wavelength range 476-684\,nm. We stacked several spectra in order to increment the S/N.  The final stacked S/N is between 25-30 at 650\,nm.  

Data reduction was performed using the ESO CPL based FLAMES/UVES Pipeline version 5.3.0\footnote{\url{http://www.eso.org/sci/software/pipelines/}} for extracting the individual fibre spectra. Data reduction includes bias subtraction, flat-field correction, wavelength calibration, and spectral rectification.

We subtracted the sky using the Sarith package in IRAF and measured radial velocities using the FXCOR package in IRAF and a synthetic spectrum as a template.  The mean heliocentric radial velocity value for our seven targets is  -71.20 $\pm$ 5.49 km $s^{-1}$. Our velocity dispersion is 14.54 $\pm$3.90 km $s^{-1}$. NGC 6440 is a massive  GC, with a mass of $5.72x10^{5}$M$\odot$ \citep{gnedin1997},  so it is expected to have a high velocity dispersion.  Using  observational data and  dynamical models \citet{gnedin02}   calculated the velocity dispersion  in  a large sample of  GCs and found in NGC 6440 a velocity dispersion of 18.2 km$s^{-1}$ at the cluster half-mass radius, one of the larger velocity dispersion among galactic GCs\footnote{http://dept.astro.lsa.umich.edu/$\sim$ognedin/gc/vesc.dat}.  In addition our sample is   concentrated toward the center of the  cluster,  therefore our measured velocity dispersion could be somewhat higher than the global value. On the other hand, as we only have seven stars the uncertainty of the dispersion is large. Our value is in agreement with the dispersion found by OR08 of 10 km$s^{-1}$. Moreover, \citet{zoccali14}  determined  the velocity  of   the bulge field stars using GIRAFFE  spectra. Along the line of sight to NGC 6440   the velocity of the field stars is  $\sim$30 km $s^{-1}$ with a dispersion of $\sim$80 km $s^{-1}$, see figure 10 from  \citet{zoccali14}. This  average  field star velocity is very different from the velocity of our sample. Finally , although  stars one  and three show the most extreme radial velocities of our sample, their chemical patterns (alpha element, iron peak elements, heavy elements) show good agreement with the rest of the sample, with a  low scatter. All these factors suggest a high probability that all of our  sample are indeed  members of NGC 6440.

The mean radial velocity is in excellent agreement with the values in the literature: \citet{saviane12} with the same sample (less  star $\#1_{-}393$) found a value of -76 $\pm$ 4 km $s^{-1}$ and \citet[2010 edition]{harris96} quotes a value of -76.6 $\pm$ 2.7 km $s^{-1}$. Table \ref{param} lists the basic parameters of the selected stars: ID, the J2000 coordinates (RA and Dec), J, H, $K_{s}$  magnitudes  from  VVV PSF photometry, calibrated on the system of 2MASS \citep{mauro14,cohen17}, heliocentric radial velocity, Teff, log(g), micro-turbulent velocity ($v_{t}$) and metallicity from our study. In addition, Table \ref{iron-abun} shows  the metallicity values from \citet{saviane12}, \citet[hereafter M14]{mauro14} and \citet{dias16a}. The determination of the atmospheric parameters is discussed in the next section. 

%__________________________________________________________________

\begin{table*}
\caption{Parameters of the observed stars.}
\label{param} \centering 
\begin{tabular}{ c c c c c c c c c c c c}
\hline 
{\small{}ID} & {\small{}Ra} & {\small{}DEC } & {\small{}J } & {\small{}H } & {\small{}K$_{s}$} & {\small{}RV$_{H}$ } & {\small{}T$_{eff}$} & {\small{}log(g)} & {\small{}{[}Fe/H{]} } & $v_{t}$\tabularnewline

 & {\small{}(h:m:s)} & {\small{}($\,^{\circ}{\rm }$:$^{\prime}$:$^{\prime\prime}$ )} & {\small{}(mag)} & {\small{}(mag)} & {\small{}(mag) } & {\tiny{}(km $s^{-1}$)} & {\small{}{[}K{]} } &  & dex & {\small{}{[}km/s{]} } \tabularnewline
\hline 
1  & 17:48:56.24 & 20:20:54.20 & 11.83 & 10.66 & 10.28  & -89.59$\pm$0.38 & 4251  & 1.92  & -0.65  & 1.76    \tabularnewline
2  & 17:48:54.47 & 20:22:41.92 & 13.51 & 12.54 & 12.27 & -61.29$\pm$0.30 & 4655  & 2.31  & -0.41  & 1.96  \tabularnewline
3  & 17:48:53.78 & 20:23:28.50 & 12.71 & 11.66 & 11.35 & -50.10$\pm$0.41 & 4429  & 2.40  & -0.51  & 2.08  \tabularnewline
4  & 17:48:55.66 & 20:21:31.50 & 12.61 & 11.48 & 11.18 & -87.16$\pm$0.49 & 4454  & 2.60  & -0.51  & 1.94   \tabularnewline
5  & 17:48:52.63 & 20:23:06.61 & 10.89 & 9.75  & 9.40 & -61.42$\pm$0.55   & 4342  & 2.50  & -0.47  & 1.56  \tabularnewline
6  & 17:48:55.45 & 20:22:19.20 & 12.95 & 11.84 & 11.55 & -71.55$\pm$0.84 & 4435  & 2.46  & -0.45  & 1.85  \tabularnewline
7  & 17:48:50.87 & 20:21:15.98 & 12.76 & 11.70 & 11.42 & -77.28$\pm$0.51 & 4450  & 2.29  & -0.52  & 2.07 \tabularnewline
\hline 
\end{tabular} 
\end{table*}

%__________________________________________________________________

\begin{table}
\caption{Iron abundances from different authors.}
\label {iron-abun}
\centering
\begin{tabular}{ c c c c c}

\hline 
\hline
ID. & [Fe/H]$_{this\_work}$ & [Fe/H]$_{S12}$ &{[}Fe/H{]}$_{M14}$&[Fe/H]$_{D16}$\\
 &  &  & & \\
\hline

1	&-0.65$\pm$0.05&-0.06$\pm$0.16&-0.37$\pm$0.14&-0.32$\pm$0.16\\
2	&-0.41$\pm$0.05&-0.07$\pm$0.16&-0.14$\pm$0.14&-0.40$\pm$0.20\\
3	&-0.51$\pm$0.05&-0.50$\pm$0.16&-0.57$\pm$0.14&-0.21$\pm$0.09\\
4	&-0.51$\pm$0.05&-0.28$\pm$0.16&-0.46$\pm$0.14&-0.15$\pm$0.16\\
5	&-0.47$\pm$0.05&-0.43$\pm$0.16&-0.52$\pm$0.14&-0.55$\pm$0.16\\
6	&-0.45$\pm$0.05&-0.12$\pm$0.16&-0.33$\pm$0.14&-\\
7	&-0.52$\pm$0.05&-0.34$\pm$0.16&-0.45$\pm$0.14&-\\

\hline
\end{tabular}
\tablefoot{S12: \citet{saviane12}; M14: \citet{mauro14}; D16: \citet{dias16a}\\

}
\end{table}
%__________________________________________________________________

\section{Atmospheric Parameters and Abundances}

The analysis of the data was performed using  the local thermodynamic equilibrium (LTE) program MOOG \citep{sneden73}.

Atmospheric models were calculated using ATLAS9 \citep{kurucz70} and  the line list for the chemical analysis is the same described in previous papers \citep[e.g.][]{villanova11}. Teff, $v_{t}$, and log(g) were adjusted and new atmospheric models calculated iteratively in order to remove trends in excitation potential and equivalent width vs. abundance for Teff and $v_{t}$ respectively, and to satisfy the ionization equilibrium for log(g). FeI and FeII were used for this latter purpose. The [Fe/H] value of the model was changed at each iteration according to the output of the abundance analysis. In  Fig. \ref{iso} we show the good agreement between our stellar parameters and an isochrone of similar metallicity as we derive ({[}Fe/H{]}=-0.5 dex) and age of 13 Gyr \citep{dotter08}.

The reddening in this cluster is a significant factor to be addressed because of its position inside the Galactic bulge. The mean color excess in the Harris catalog  is E(B-V)=1.07 \citep[2010 edition]{harris96}, similar to that of \citet{valenti04}, E(B-V) = 1.15. Additionally there is a strong and complex  differential reddening (Figure \ref{reddening}). In our case the stellar parameters were found directly from the spectra, as explained above, so our measurement of abundances  is not affected by the effects of reddening.

%----------------------------------------------------------------
  
 \begin{figure}
\centering
 \includegraphics[width=3.7in,height=4.0in]{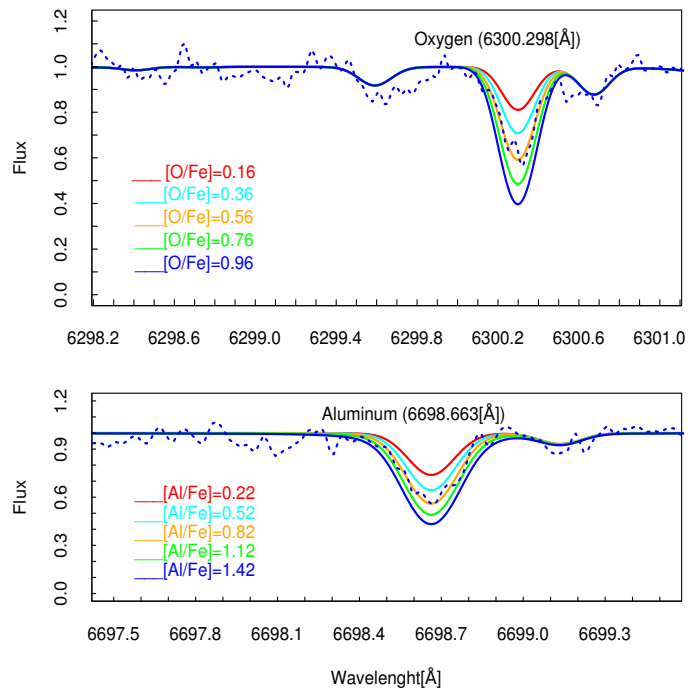}
 \caption{Spectrum synthesis fits for  Oxygen (star \#3)   and Aluminum (star \#4) lines  respectively. The dotted line is the observed spectrum and the solid color lines show the synthesised spectra corresponding to different abundances. }
 \label{synth}
 \end{figure}

%----------------------------------------------------------------
The chemical abundances for Ca, Ti, Fe, Co and Ni were obtained using equivalent widths (EWs) of the spectral lines; a more detailed explanation of the method we used to measure the EWs is given in \citet{marino08}. For the other elements (O, Na, Mg, Al, Si, Sc, Ni, Mn, Ba and Eu), whose lines are affected by blending, we used the spectrum-synthesis method. We calculated five synthetic spectra having different abundances for each line, and estimated the best-fitting value as the one that minimises the rms scatter. We show in Figure \ref{synth} an example of the method for two lines (Oxygen
and Aluminum). Only lines not contaminated by telluric lines were used. The adopted solar abundances we used are reported in Table \ref{abundances}.

%----------------------------------------------------------------
\begin{figure}
\centering
    \includegraphics[width=2.7in,height=3.2in]{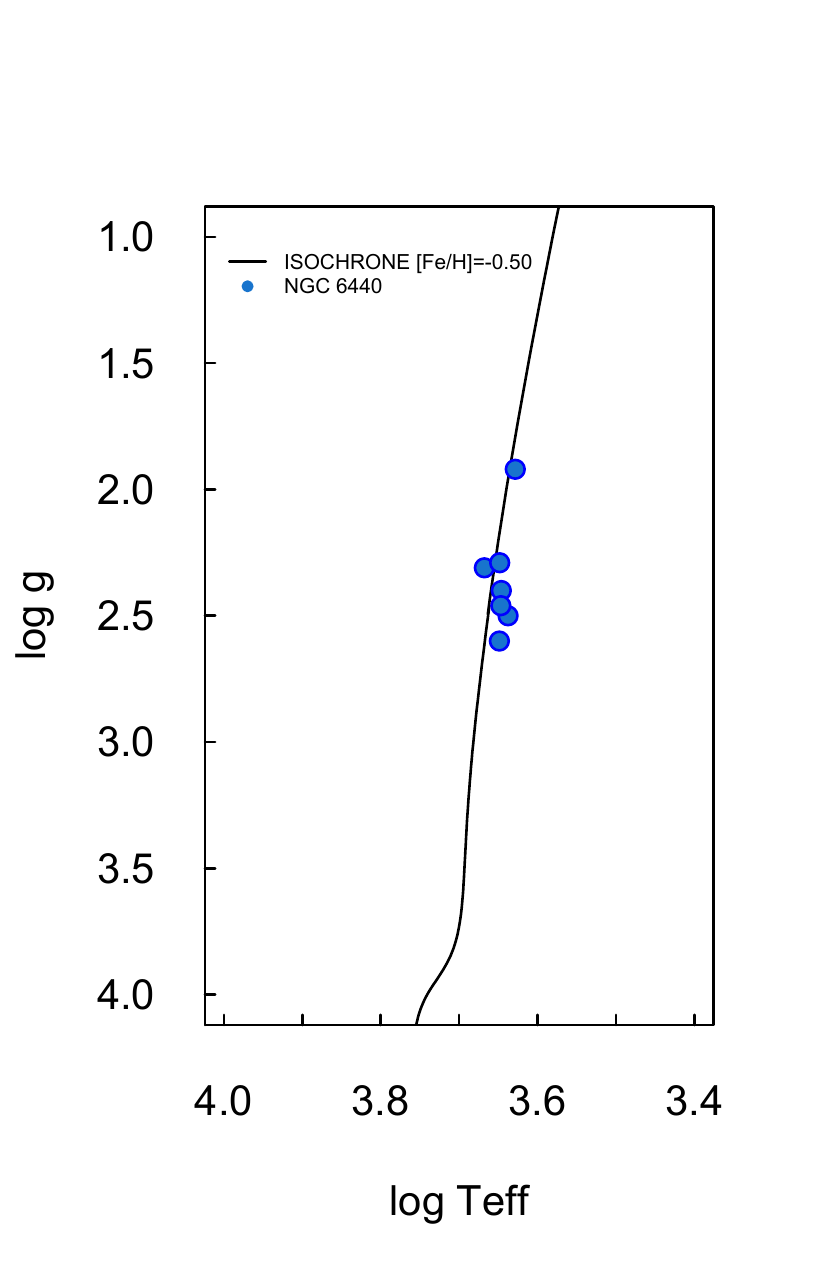}
   \caption{Log g vs log Teff for our sample of seven stars. The over-plotted isocrone has a metallicity of -0.5 dex,[$\alpha$/Fe]=+0.40 dex  and age of 13 Gyr \citep{dotter08}.}
  \label{iso}
 \end{figure}
 
%----------------------------------------------------------------
An internal error analysis was performed by varying $T_{eff}$, log(g), {[}Fe/H{]}, and $v_{t}$ and redetermining abundances of star \#3, assumed to be representative of the entire sample.
Parameters were varied by $\Delta T_{eff}=+40$ K, $\Delta$log(g)=+0.24, $\Delta${[}Fe/H{]}=+0.03 dex, and $\Delta v_{t}=+0.08$ $km$ $s^{-1}$, which we estimated as our typical internal errors.
 The amount of variation of the parameter was calculated using three stars representative of our sample (\#1,\#3, and \#2)  with  relatively low, intermediate and high effective temperature respectively, according to the procedure that was performed by \citet{marino08}, which we use in this study.

the error introduced by the uncertainty on  the  EW ($\sigma_{S/N}$) was calculated by dividing the rms scatter  by the square root of the number of the lines used for a given element and a given star. For elements whose abundance was obtained by spectrum-synthesis, the error is the output of the fitting procedure.

Finally the error for each [X/Fe] ratio as a result of  uncertainties in atmospheric parameters and $\sigma_{S/N}$ are listed in Table \ref{error}. The total internal error ($\sigma_{tot}$) is given by:

\begin{center}
\begin{equation}
\sigma_{tot}=\sqrt{\sigma^{2}_{T_{eff}}+\sigma^{2}_{log(g)}+\sigma^{2}_{v_{t}}+\sigma^{2}_{[Fe/H]}+\sigma^{2}_{S/N}}
\end{equation}
\end{center}

In Table \ref{error} we compare the total internal error for each element with the observed error (standard deviation of the sample).

%----------------------------------------------------------------
     \begin{figure}
\centering
  \includegraphics[width=3.25in,height=3.45in]{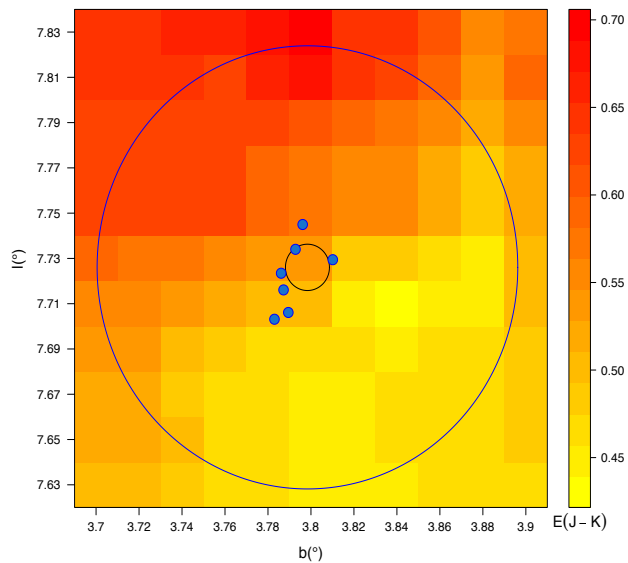}
  \caption{VVV reddening map for  NGC 6440 \citep{gonzalez12}. The large blue circle is the tidal radius  and small circle is the half light radius. The blue filled  circles are our observed UVES sample.}
  \label{reddening}
 \end{figure}
 
 %----------------------------------------------------------------
%______________________________________________________________
 
\begin{table*}
\caption{Abundances of the observed stars.}
\label {abundances}
\centering
\begin{tabular}{ l c  c c  c  c c  c  c c c  }
\hline 
\hline

El. & 1 & 2  & 3 & 4  & 5 & 6 & 7  & Cluster \tablefootmark{1} &Sun   &$N^{\circ}$ Lines\\

\hline 
$ [$O/Fe$] $   &   0.57  &	0.43  &	0.57  &	0.52&	0.57   &	0.67   &	0.30     & +0.52$\pm$0.05   & 8.83 & 1\\
&$\pm$0.04&$\pm$0.05&$\pm$0.05&$\pm$0.04&$\pm$0.05&$\pm$0.06&$\pm$0.04&\\

$ [$Na/Fe$]_{NLTE} $  &	 0.15  &	0.21 &	0.48  &	0.44 &	0.53  &	0.85     &	1.05  & +0.53$\pm$0.12   & 6.32&2\\
&$\pm$0.06&$\pm$0.09&$\pm$0.09&$\pm$0.08&$\pm$0.07&$\pm$0.07&$\pm$0.08&\\

$ [$Mg/Fe$] $  &	 0.41  &	0.44   &	0.39   &	0.37  &	0.24   &	0.47   &	0.29     & +0.37$\pm$0.03   & 7.56&3\\
&$\pm$0.05&$\pm$0.08&$\pm$0.05&$\pm$0.06&$\pm$0.04&$\pm$0.06&$\pm$0.06&\\

$ [$Al/Fe$] $  &	 0.58  &	0.32   &	0.34   &	0.81  &	0.64   &	0.56   &	0.69    & +0.56$\pm$0.07  & 6.43&2\\
&$\pm$ 0.04&$\pm$0.04&$\pm$0.04&$\pm$0.04&$\pm$0.04&$\pm$0.03&$\pm$0.04  \\

$ [$Si/Fe$] $  &	 -0.05   &	0.67   &	0.13   &	0.13  &	0.02   &	0.24   &	0.26    & +0.20$\pm$0.09   & 7.61&2\\
&$\pm$0.07&$\pm$0.08&$\pm$0.08&$\pm$0.09&$\pm$0.06&$\pm$0.09&$\pm$0.07&\\

$ [$Ca/Fe$] $  &	 0.69   &	0.19   &	0.14   &	0.35 &	0.38   &	0.25   &	0.41     & +0.34$\pm$0.07   & 6.39&4\\
&$\pm$0.07&$\pm$0.05&$\pm$0.07&$\pm$0.08&$\pm$0.08&$\pm$0.08&$\pm$0.09&\\

$ [$Sc/Fe$] $  &	 0.25  &	0.10  &	0.36 &	0.28 &	0.34   &	0.19   &	0.18    & +0.24$\pm$0.03   & 3.12&2\\
&$\pm$0.07&$\pm$0.09&$\pm$0.09&$\pm$0.09&$\pm$0.06&$\pm$0.09&$\pm$0.07&\\

$ [$Ti/Fe$] $\tablefootmark{2}  &0.41  &	0.28   &	0.49  &	0.30 &	0.60   &	0.48     &	0.36  & +0.42$\pm$0.04  & 4.94&12\\
&$\pm$0.05&$\pm$0.07&$\pm$0.07&$\pm$0.06&$\pm$0.07&$\pm$0.06&$\pm$0.07&\\

$ [$Mn/Fe$] $   &	 -0.08&	-0.10  &	-0.12  &	-0.01 &	0.11  &	0.13  &	-0.01  & -0.01$\pm$0.04   & 5.37&1\\
&$\pm$0.08&$\pm$0.07&$\pm$0.08&$\pm$0.09&$\pm$0.08&$\pm$0.07&$\pm$0.10&\\

$ [$Fe/H$] $   &	 -0.65&	-0.41  &	-0.51  &	-0.51 &	-0.47  &	-0.45  &	-0.52  & -0.50$\pm$0.03   & 7.50&$\sim$100 \\
&$\pm$0.03&$\pm$0.02&$\pm$0.02&$\pm$0.03&$\pm$0.04&$\pm$0.03&$\pm$0.03&\\

$ [$Co/Fe$] $  &	 0.50   &	0.40   &	0.42   &	0.51  &	0.58   &	0.38   &	0.49    &+ 0.42$\pm$0.03   & 4.93&10\\
&$\pm$0.06&$\pm$0.07&$\pm$0.07&$\pm$0.07&$\pm$0.08&$\pm$0.06&$\pm$0.08&\\

$ [$Ni/Fe$] $  &	  0.15  &	-0.08  &	-0.05  &	-0.07 &	0.19 &	-0.15  &	0.06  &+0.01$\pm$0.05  & 6.26&7\\
&$\pm$0.06&$\pm$0.08&$\pm$0.08&$\pm$0.07&$\pm$0.08&$\pm$0.06&$\pm$0.08&\\

$ [$Ba/Fe$] $  &	 -0.21  &	-0.29 & -0.12 &   -0.25  &	-0.01  &	-0.09  & -0.19   & -0.17$\pm$0.05  & 2.34&1\\
&$\pm$0.07&$\pm$0.09&$\pm$0.06&$\pm$0.07&$\pm$0.07&$\pm$0.06&$\pm$0.07&\\

$ [$Eu/Fe$] $  &	 0.45  &	0.39  &	0.38  &0.52  &	0.51  &	0.50  &	0.40   & +0.45$\pm$0.02 & 0.52&1\\
&$\pm$0.06&$\pm$0.05&$\pm$0.06&$\pm$0.05&$\pm$0.06&$\pm$0.05&$\pm$0.06&\\

\hline

\end{tabular}
\tablefoot{
Columns 2-8: abundances of the observed stars. Column 9: mean abundance for the cluster. Column 10: abundances adopted for the Sun in this paper. Abundances for the Sun are indicated  as log$\epsilon$(El.).Column 11: number of measured lines for each chemical element.
The errors presented for each abundance was calculated by dividing the rms scatter by the square root
of the number of the lines used for a given element and a
given star. For elements whose abundance was obtained by
spectrum-synthesis, the error is the output of the fitting procedure.\\
\tablefoottext{1}{The errors are the statistical errors obtained of the mean.}\\
  \tablefoottext{2}{ [Ti/Fe] is the average between Ti I and Ti II.}
 
}
\end{table*}

 %______________________________________________________________

\begin{table*}
\caption{Estimated errors on abundances, due to errors on atmospherics parameters and to spectral noise, compared with the observed errors.}
\label {error}
\centering
\begin{tabular}{ l c  c c  c  c c  c  c c  c }
\hline 
\hline
	ID   &  $\Delta T_{eff} =40 K $ & $ \Delta log(g)=0.24$  & $\Delta v_{t}= 0.08$	& $ \Delta [$Fe/H$]=0.03 $ & $\sigma_{S/N} $& $\sigma_{tot}$  & $\sigma_{obs}$\\		
	\hline					
	$ \Delta ([$O/Fe]$) $     &	-0.02 &	0.08   &	-0.01   &	0.00   &	0.05  &	0.10  &	0.12\\ 
	$ \Delta ([Na/Fe]) $  &	-0.05 &	0.04  &	-0.01   &	0.03   &	0.09 &	0.11 &	0.32\\ 	
	
	$ \Delta ([$Mg/Fe$]) $    &	-0.01  &	-0.02   &	-0.01   &	0.01   &	0.05  &	0.06  &	0.08\\ 	
	$ \Delta ([$Al/Fe$]) $    &	-0.01  &	-0.03  &	0.00   &	0.01   &	0.04  &	0.05  &	0.18\\ 	
	
	$ \Delta ([$Si/Fe$]) $    &	0.03   &	-0.04  &	0.01   &	0.02   &	0.08  &	0.10  &	0.23\\ 
			
	$ \Delta ([$Ca/Fe$]) $    &	-0.04   &	-0.01  &	0.00  &	0.01  &	0.07  &	0.08 &	0.18\\

	$ \Delta ([$Sc/Fe$]) $    &	0.00   &	0.04  &	0.01  &	-0.01   &	0.09&	0.10 &	0.09\\ 	
	
	$ \Delta ([$Ti/Fe$]) $  &	-0.04   &	0.00 &	0.00   &	0.02  &	0.07 &	0.08  &	0.11\\ 	
	
  	  $ \Delta ([$Mn/Fe$]) $    &	0.01  &	0.09   &	0.01   &	0.01   &	0.08  &	0.12 &	0.09\\ 	

	$ \Delta ([$Fe/H$]) $     &	-0.01   &	0.03   &	0.03   &	0.02   &	0.02  &	0.05  &	0.08\\ 
    
	$ \Delta ([$Co/Fe$]) $    &	-0.02   &	0.05   &	0.00  &	0.02  &	0.07  &	0.09  &	0.07\\ 	
	$ \Delta ([$Ni/Fe$]) $  &	-0.03  &	-0.04   &	0.05   &	0.03   &	0.08  &	0.11 &	0.13\\ 	
	$ \Delta ([$Ba/Fe$]) $    &	-0.02  &	0.08   &	0.00   &	0.07   &	0.06  &	0.12  &	0.10\\ 	
	$ \Delta ([$Eu/Fe$]) $    &	0.01  &	0.09   &	0.01   &	0.01   &	0.06  &	0.11 &	0.06\\

\hline

\end{tabular}
\end{table*}
%______________________________________________________________

\section {Results}
In this sections, we will discuss in detail our results. In addition, we compare them with the literature, and also do a general comparison  with bulge globular clusters analysed up to this moment.

 \subsection{Iron}
We found a mean [Fe/H] value for the cluster of:

\begin {center}
\vspace{0.3cm}
[Fe/H]=$-0.50\pm0.03$ dex
\vspace{0.3cm}
\end {center}

The observed scatter is consistent with that expected solely from errors and thus we find no evidence for any intrinsic Fe abundance spread. However, star \#1 of our sample has a difference of 0.12 dex compared to the average: in order to decide whether this is a peculiar star, or if there is a real intrinsic dispersion, a larger sample would be needed. Hereafter we have highlighted this star in most plots,
to check for any other peculiarity. We now discuss how our result compares to the literature. 

 \citet{saviane12}, who selected the targets we analyse in this paper, measured CaII triplet equivalent widths of all stars. They applied the metallicity scale from \cite{carretta09c} and found an average of [Fe/H]=-0.25$\pm$0.16 for eight stars in NGC~6440. In Table \ref{iron-abun} we list the metallicities of  \citet{saviane12} for our stars. If we use the new metallicity scale of \cite{dias16a,dias16b} the average is [Fe/H]=-0.28$\pm$0.14. This is compatible with our results within 1.5-$\sigma$. No evidence of intrinsic Fe abundance spread was found, in agreement with our findings.

 \citet{mauro14} used the CaII triplet equivalent widths from \citet{saviane12} and corrected them by gravity and temperature effects using infrared magnitudes, instead of the traditional V magnitudes used by \citet{saviane12}. 

 They found an average metallicity of [Fe/H]=-0.38$\pm$0.14 
 
using the scale from \citet[][see Table \ref{iron-abun}]{carretta09c}, which becomes [Fe/H]=-0.29$\pm$0.14 if we adopt the new metallicity scale of \cite{dias16a,dias16b}.
We do not confirm the possible metallicity spread suggested by \citet{mauro14}.

 \citet{dias15,dias16a} analysed the low-resolution optical spectra of the same stars selected by \citet[][see Table \ref{iron-abun}]{saviane12}. They applied full-spectrum fitting against a synthetic and an empirical library. Their results agree with homogeneous high-resolution spectroscopic homogeneous results within 0.08~dex   and they defined a new metallicity scale based on their sample of 51 GGCs. They found an average metallicity for NGC~6440 of {[}Fe/H{]}=-0.24$\pm$0.05~dex, based on the same seven stars.

 The three  published results above from CaII triplet and optical spectra of the same stars agree very well. Nevertheless we analysed the same stars with high-resolution spectroscopy and our result is about 0.2~dex more metal-poor. The typical error for the metallicity of a single star in the three analysis above is about 0.15~dex, therefore, the 0.2~dex different may be explained by the error bars. We have added the metallicities for individual stars from the three analysis in  Tab. \ref{iron-abun} for reference.

OR08 derived stellar parameters and abundances for Fe, O, Ca, Si, Mg, Al and C in NGC 6440 using high resolution infrared spectra. Although we have no common star to make a direct comparison, the average of the sample from OR08 (10 RGB stars) for iron is {[}Fe/H{]}=-0.54 dex with a $\sigma$=0.06, showing excellent  agreement with our results.

%----------------------------------------------------------------

 The field of NGC~6440 is greatly affected by a large differential reddening (see Figure \ref{reddening}). \citet{valenti04}  used NIR photometry (J, H and K bands) and also found an iron abundance of  {[}Fe/H{]}=-0.49 dex. \citet{cohen17} using a near-IR photometric metallicity calibration from VVV PSF photometry found an iron abundance of  {[}Fe/H{]}=-0.46 dex. Both studies using  photometry show good  agreement with ours. 
 Finally, the  metallicity quoted by \citet[2010 edition]{harris96} of -0.36 dex is an average of the values above, mainly based on photometric results.
 
As it stands, there are some results showing an average metallicity of [Fe/H]$\sim$-0.3 for NGC~6440, and others showing [Fe/H]$\sim$-0.5 in agreement with our result. However, the two most reliable studies - based on high resolution optical and near-IR spectra - are in very good agreement and strongly suggest the metallicity is close to -0.5. More bulge clusters need to be analysed homogeneously  in order to compare the metallicity scale.

%______________________________________________________________

\subsection{Iron-peak elements}

The abundance pattern of iron-peak elements may be of great help in identifying the  ISM polluters from  among the variety of possible sources. In particular, discriminating between SN and other polluters 
can give clues about  the chemical evolution of the ISM from which the cluster has formed.

\begin{figure}
\centering
  \includegraphics[width=3.5in,height=3.8in]{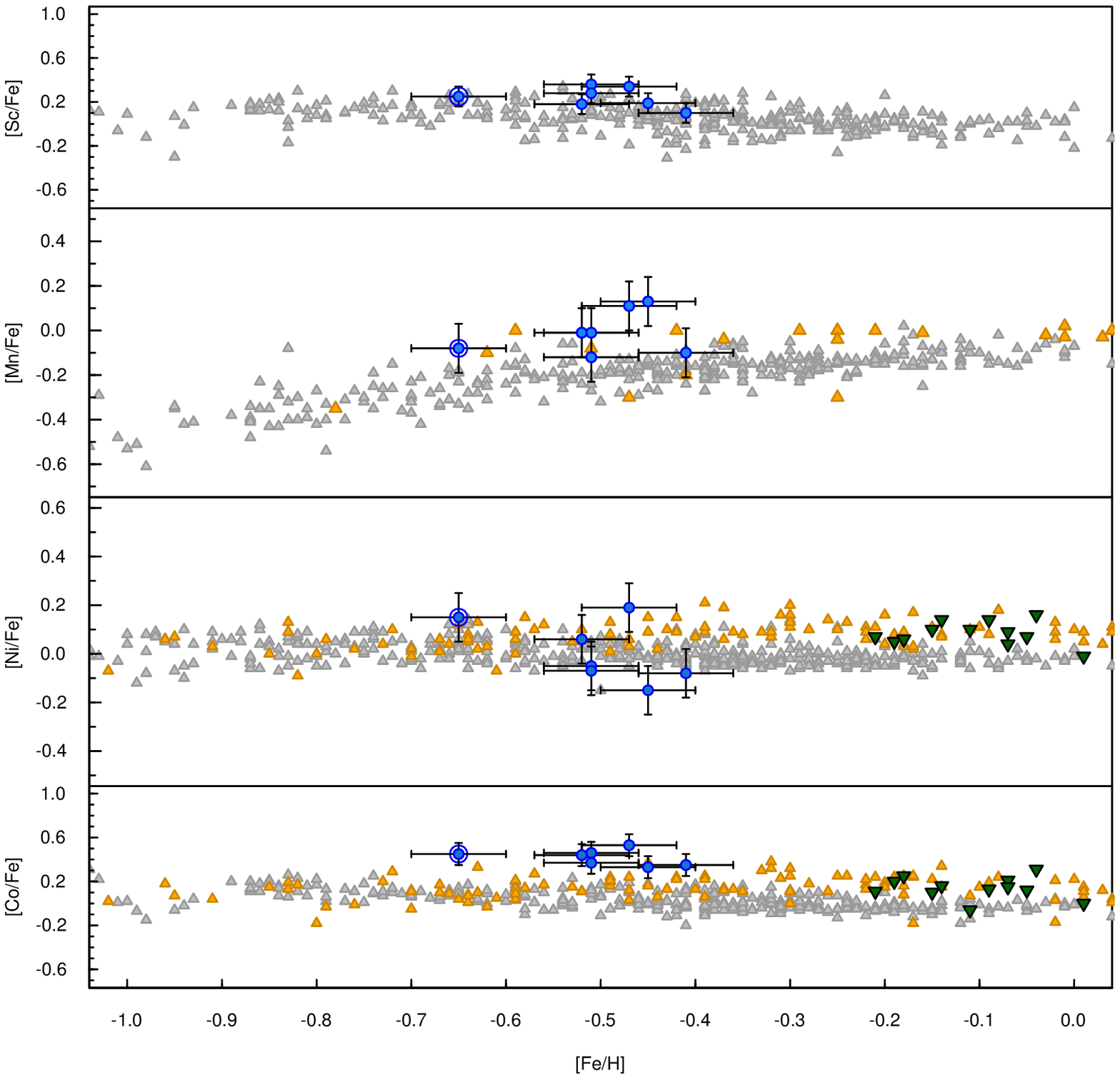}
  \caption{[Sc,Mn,Ni,Co/Fe] vs [Fe/H]. Filled blue circles  with error bars are our data for NGC 6440, filled orange triangles: Bulge field stars \citep{barbuy13,johnson14}, filled gray triangles: halo and disk  stars \citep{fulbright00,francois07,reddy03,reddy06}, filled dark green triangles:  NGC 6553 \citep{johnson14}.}
  \label{iron-ele}
 \end{figure}

In this study we have measured the abundances of five iron peak elements: Fe, Sc, Mn, Co and Ni (see Table \ref{abundances}). We have discussed extensively about Fe in the previous section.\\
Figure \ref{iron-ele} shows the other iron-peak elements {[}Sc, Mn, Co and Ni/Fe{]} vs. {[}Fe/H{]} compared with abundances for halo, disk and bulge field stars, including some Bulge GCs. We have found  super-solar values for  Sc and Co. 

 %----------------------------------------------------------------
 %----------------------------------------------------------------
 Other authors have found super-solar abundances for iron-peak elements in bulge field stars: for example \citet{mcwilliam94} found an average for {[}Co/Fe{]} = 0.28 dex and {[}Sc/Fe{]} = 0.33 dex, which are similar to our values after taking into account our errors (0.09 dex and 0.10 dex, respectively). Their solar value for nickel is also in agreement with our result. In a similar way, \citet{johnson14} found an enhanced value of the cobalt abundance of bulge field stars, although not as large as ours ({[}Co/Fe{]} = 0.14 dex). Although their absolute abundances are somewhat different from  ours, \citet{johnson14} also found a larger enhancement of cobalt with respect to nickel (see figure \ref{iron-ele} and Table \ref{abundances}).
 
 Next we compare  the iron-peaks elements of NGC 6440  with NGC 6553, which is  another Bulge GCs, part of the same study of  \citet{johnson14}. We see similar patterns, although they do find for cobalt an average of [Co/Fe] = 0.22 dex and nickel an average of [Ni/Fe] = 0.10 dex, slightly low  and slightly high compared  to  our values, respectively.  However, if we take into account the difference in metallicity, the chemical patterns of iron-peak elements  essentially follow a comparable  pattern, where the enhancement of cobalt is larger with respect to nickel (see figure \ref{iron-ele} and Table \ref{abundances}).

Some studies suggest that significant enhancements of Co and Sc, as observed in this study,  can be due to a very light neutron exposure of the atmospheric materials \citep{smith87,mcwilliam94}. Additionally, both elements Co and Sc could be produced by SN explosion.

Coming to manganese (Mn), \citet{cescutti08} have reproduced the behavior of Mn vs Fe for the bulge of the Milky Way taking into account the metallicity dependence of yields for SNeII and SNe Ia. Their models are in agreement with our results, indicating a strong   contribution by SN explosion.

In summary, the high abundance of the iron-peak elements in NGC 6440, together with the fact that it does not show a significant iron spread,  indicates that NGC 6440 formed in an environment with a high abundance of these elements, which are produced  mainly by  SNeIa and SNeII.

%______________________________________________________________
\subsection{$\alpha$ elements}
\label{alpha}

The chemical abundances for the $\alpha$ elements O, Mg, Si, Ca, and Ti listed in Table \ref{abundances} are all significantly  overabundant relative to solar scales. If we use Mg, Si, Ca, and Ti to estimate a mean $\alpha$-element value (O will be treated separately) we obtain:
\begin{center}
\vspace{0.2cm}
 {[}$\alpha$/Fe{]}=$0.35\pm0.06$ \vspace{0.3cm}
 \par\end{center}
OR08 also measured alpha elements for NGC 6440. These are, in general, in good agreement with our values (see Table \ref{ces-ori} and Figure \ref{comp-OR08}). The most significant discrepancies are for Oxygen and Aluminum. These elements will be discussed separately below in the subsection on O-Na and Mg-Al, respectively.

\begin{figure}
\centering
  \includegraphics[width=3.5in,height=3.5in]{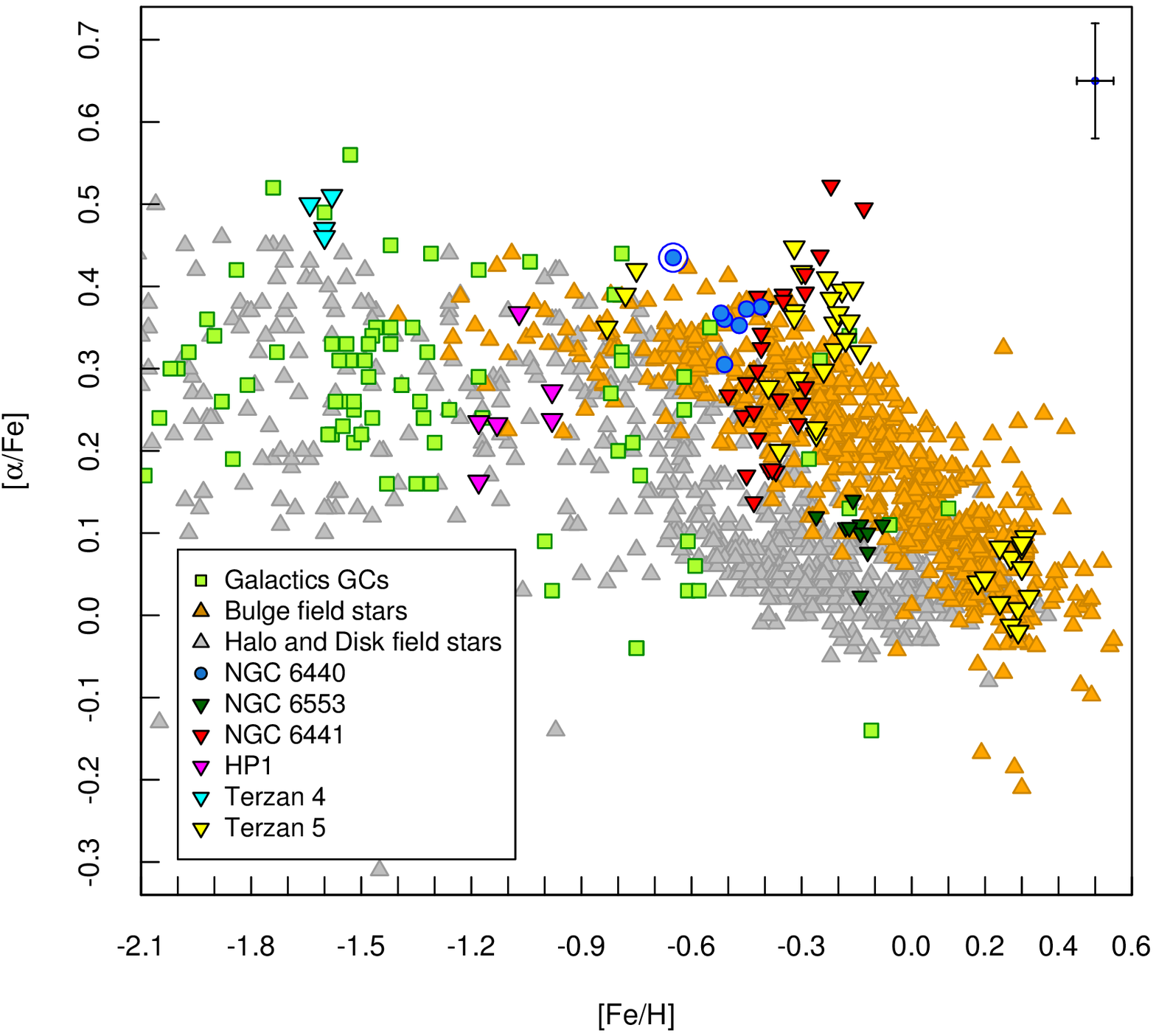}
  \caption{[alpha/Fe] vs [Fe/H].Filled blue circles are our data for NGC 6440, filled red triangles :  NGC6441 \citep{gratton06,gratton07}, filled yellow triangles: Terzan 5 \citep{origlia11,origlia13}, filled cyan triangles: Terzan 4\citep{origlia04}, filled dark green triangles: NGC 6553 \citep{tang17}, filled magenta triangles: HP1 \citep{barbuy16}, filled orange triangles: bulge field stars\citep{gonzalez12}, filled green square: GCs from \citet{pritzl05},  filled grey triangles : Halo and Disk fields stars \citep{venn04}.}
  \label{alpha1}
 \end{figure}

\begin{figure*}
\centering
  \includegraphics[width=6.1in,height=6.6in]{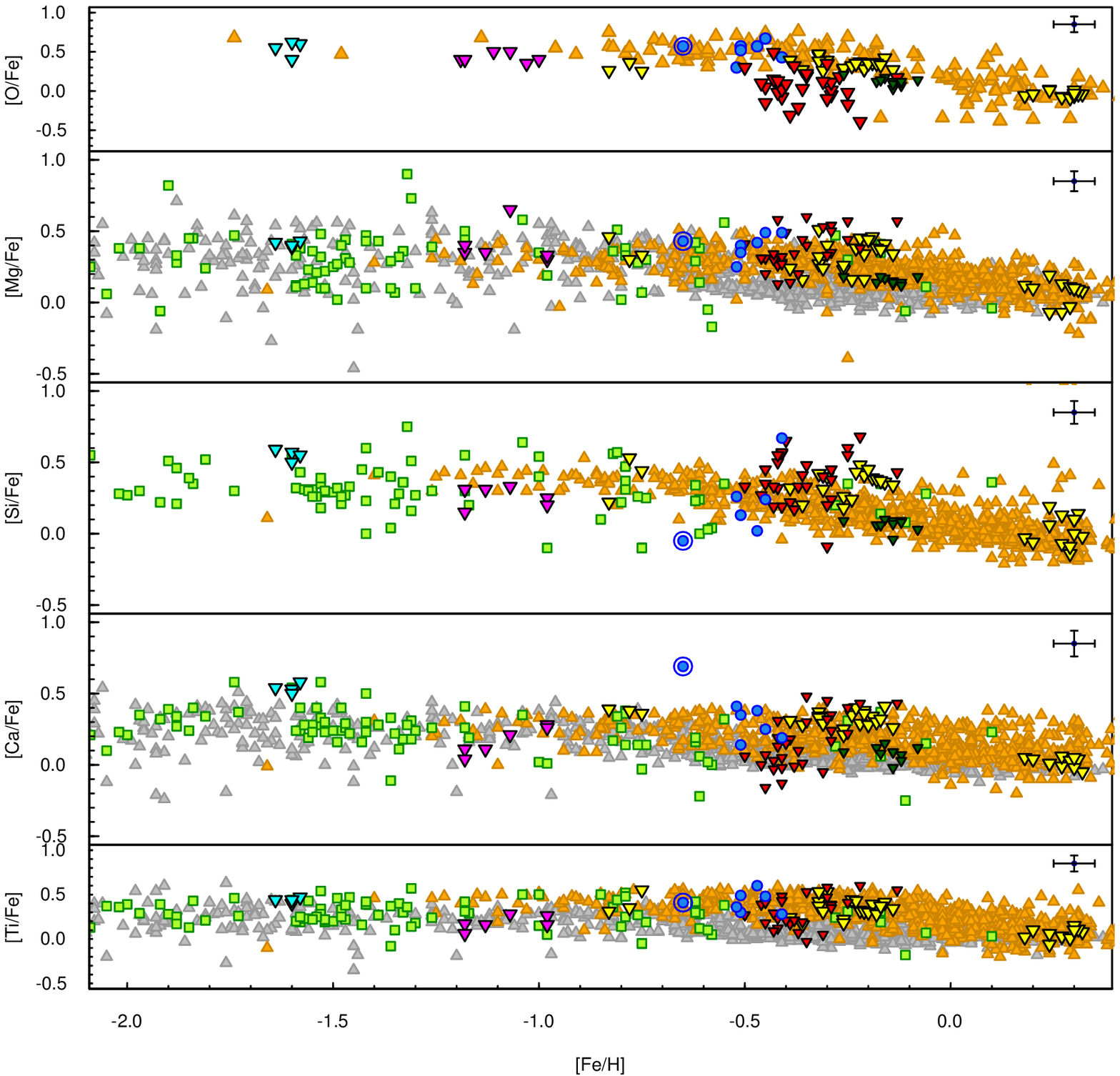}
  \caption{$[$O/Fe$]$,[Mg/Fe], [Si/Fe], [Ca/Fe], [Ti/Fe] vs [Fe/H].  Filled blue circles are our data for NGC 6440, filled red triangles :  NGC6441 \citep{gratton06,gratton07}, filled yellow triangles: Terzan 5 \citep{origlia11,origlia13}, filled cyan triangles: Terzan 4\citep{origlia04}, filled dark green triangles: NGC 6553 \citep{tang17}, filled magenta triangles: HP1 \citep{barbuy16}, filled orange triangles: bulge field stars\citep{gonzalez12}, filled green square: GCs from \citet{pritzl05},  filled grey triangles : Halo and Disk fields stars \citep{venn04}.}

  \label{alpha2}
 \end{figure*}

\begin{figure}
%\centering
  \includegraphics[width=3.1in,height=3.8in]{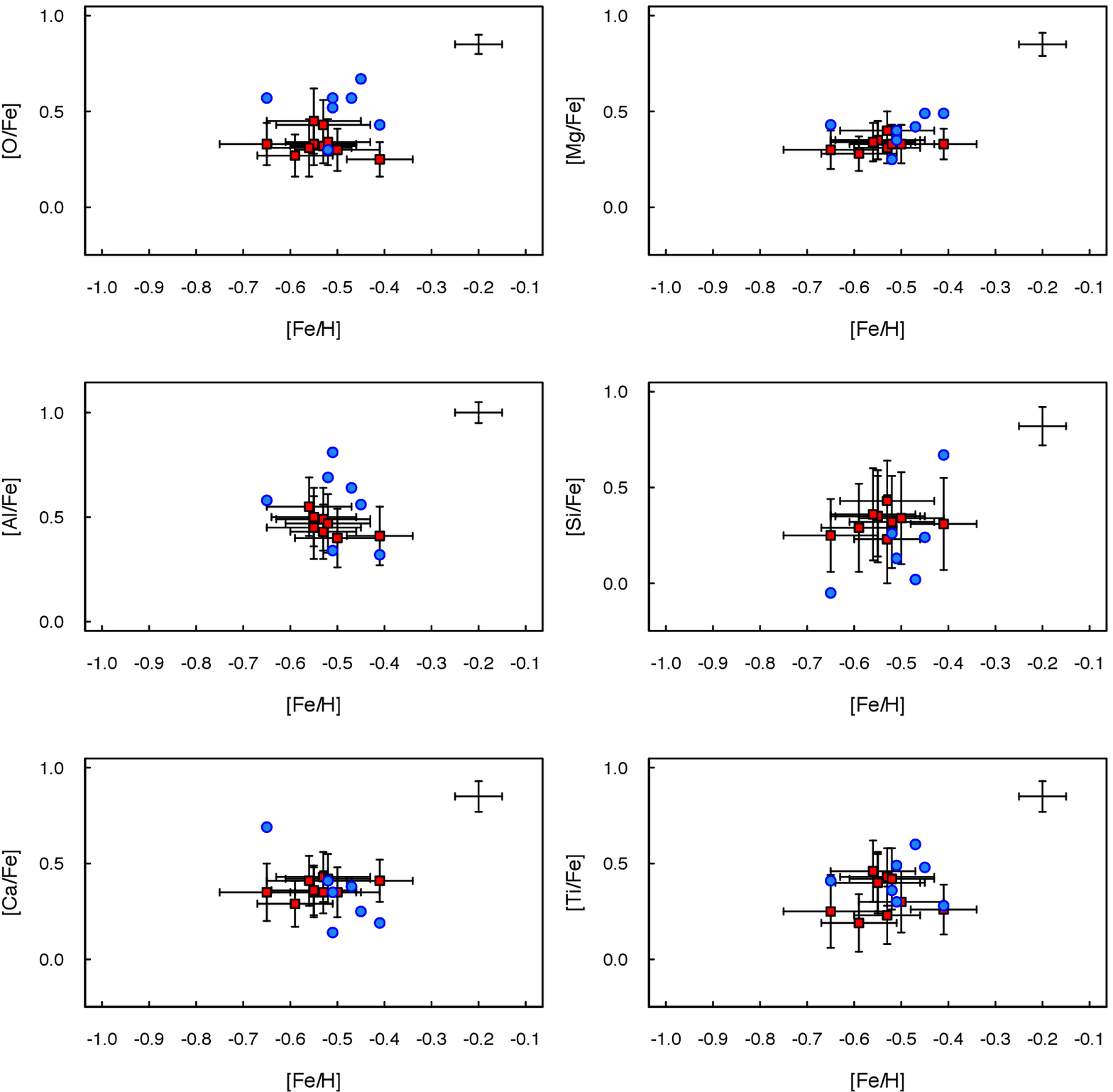}
  \caption{Comparison between this study (blue filled circles) with OR08 (red filled squares) for NGC 6440.}
  \label{comp-OR08}
 \end{figure}

In  Figures  \ref{alpha1} and \ref{alpha2} we note that NGC 6440 follows the trend of bulge field stars for the alpha elements, in accordance also with other bulge GCs that follow the same  trend (NGC 6441, Terzan 5, Terzan 4, HP1 , NGC 6553). 

 %----------------------------------------------------------------

%----------------------------------------------------------------

%----------------------------------------------------------------

%-------------------
Analysing the  total observational error and the actual  dispersion (Table \ref{error}), which tell us whether for some elements there might be an internal spread, we note that for the alpha elements: O, Mg and Ti, there is a good agreement between total and observed dispersion, however for  both Si and Ca  the actual spread is significantly larger than expected due to errors. Unlike the light elements, we do not expect Si and Ca to show an intrinsic variation. Note that oxygen doesn't show significant scatter (see section 1, 4.4 and 4.5). It is necessary, therefore, to corroborate this behavior using a  large sample of stars.

When comparing these results for NGC 6440 with other bulge GCs,  we noticed that NGC 6441, NGC 6553, Terzan 5 and HP1 also  show some spread in  their $\alpha$-elements, on the other hand the scatter in the alpha element abundances for bulge field stars is evident.

Many authors agree that the bulge formation was rapid, as shown by the enhanced alpha plateau against {[}Fe/H{]} in bulge field stars \citep[e.g.][]{ballero07,cescutti11}. This enhancement in alpha elements is produced by massive stars exploding as SNe II at early epochs, producing super-solar alpha abundances. Then,  this plateau eventually turns down as metallicity increases, due to the onset of SN Ia yielding mainly iron-peak elements without alphas. 

For higher SF rates, more iron is produced before SNe Ia start changing the composition of the ISM, and the knee  occurs at  higher metallicities. The location of the knee for stars in the bulge shows that its evolution was faster than that of the halo and the disk.

In this context, we note that all alpha elements in NGC 6440 are overabundant, and especially oxygen, indicating an early enrichment by SNe II. This is in good agreement with the alpha enhancement that is shown by bulge stars. The other bulge GCs, such as HP1, NGC 6441, NGC 6553, and Terzan 5, also follow the trend of the bulge.

Analyzing the [O/Fe] vs. [Fe/H], note that  the bulge GC NGC 6441 shows an Oxygen abundance generally lower than bulge field stars (with scatter), indicating a depletion of oxygen, in accordance with the model of self enrichment in GCs. In contrast, NGC 6440 follows very well the trend of the bulge with low spread.

%______________________________________________________________
\subsection{Na-O anticorrelation}

 Many Galactic GCs so far studied  shows an anticorrelation between Na and O abundances, which is the most recognized evidence for the existence  of MP in GCs \citep{carretta09a,carretta09b,gratton12}.
The only old GC that apparently does not show this anticorrelation is Ruprecht 106 \citep{villanova13}, but the origin (galactic or extragalactic) of this object is in doubt.

According to the models, GCs show this anticorrelation due to the fact that the material of some of the stars we observe has been processed through proton-capture nucleosynthesis by the CNO cycle, which depletes  oxygen, while the NeNa chain enriches Na \citep{dantona06}. It is therefore postulated that a first generation of stars is followed by a second generation which is much more Na-rich and O-poor. The Na-O anticorrelation in NGC 6440 is partly seen in our  data, but it is not so extended as commonly seen in most  galactic GCs.

Comparing the measurement dispersion of Na ($\sigma_{tot}=0.11$ dex) with the observed one ($\sigma_{obs}=0.32$ dex), there is clearly a significant intrinsic  spread, while the opposite is true for oxygen. The observed spread ($\sigma_{obs}=0.12$ dex) is consistent with measurement errors ($\sigma_{tot}=0.10$ dex), at odds with most  galactic GCs.

\begin{figure}
\centering
  \includegraphics[width=3.3in,height=3.3in]{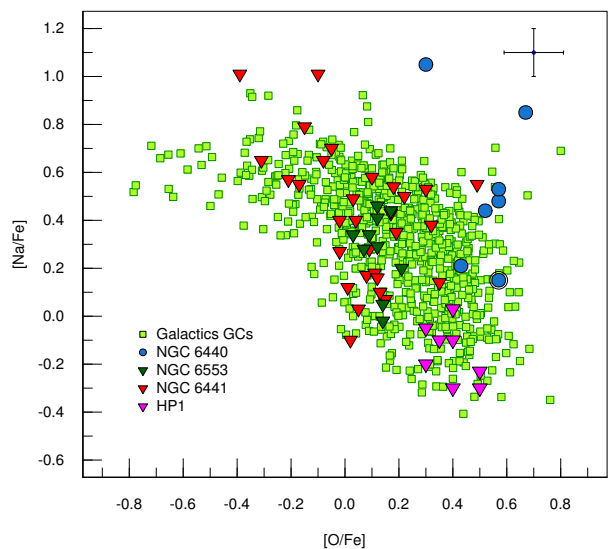}
  \caption{[O/Fe] vs [Na/Fe].
  Filled blue circles are our data for NGC 6440, filled red triangles :  NGC6441 \citep{gratton06,gratton07}, filled dark green triangles: NGC 6553 \citep{tang17}, filled magenta triangles: HP1 \citep{barbuy16},  filled green square: Galactic GCs from \citet{carretta09b}.}
  \label{o-na}
 \end{figure}

 A low intrinsic  oxygen dispersion is also found for most of the other bulge clusters. For example, OR08 measured  oxygen abundances in ten stars of NGC 6440. Although  our mean values differ by 0.19 dex,  as shown in the Table \ref{ces-ori} and Figure \ref{comp-OR08}, they  also found no significant spread in oxygen, in agreement with our finding. HP1  is another bulge cluster which shows no clear  O-Na anticorrelation (Fig. \ref{o-na}) according to \citet{barbuy16}. On the other hand, HP1 has a relatively low mass, therefore
it is likely not to show this anticorrelation \citep{barbuy16}. A low spread in oxygen was also found in NGC 6553 \citep{tang17} in a sample of 10 stars. Terzan 4 shows a similar behavior, with a  low spread in oxygen \citep{origlia04}, although their sample is only four stars. The only exception to this picture is NGC 6441, which possesses an extended O-Na anticorrelation. It should be noted that  the mass of this cluster is the largest among  bulge GCs compared in this study (Figure \ref{o-na} ), which makes the possibility of self-enhancement more likely.\\

%______________________________________________________________

\begin{table}
\caption{Mean abundance of elements for this work and OR08.}
\label {ces-ori}
\centering
\begin{tabular}{ c c c }

\hline 
\hline
el. & This work & OR08\\
\hline

$[$Fe/H$]$	&$-0.51\pm0.03 $&$-0.54\pm0.02$\\
$[$O/Fe$]$	&$0.52\pm0.05$&$0.33\pm0.02$\\
$[$Mg/Fe$]$	&$0.37\pm0.03$&$0.33\pm0.01$\\
$[$Al/Fe$]$	&$0.56\pm0.07$&$0.46\pm0.02$\\
$[$Si/Fe$]$	&$0.20\pm0.09$&$0.32\pm0.02$\\
$[$Ca/Fe$]$	&$0.34\pm0.07$&$0.37\pm0.01$\\
$[$Ti/Fe$]$	&$0.42\pm0.04$&$0.33\pm0.03$\\

\hline

\end{tabular}\\
\tablefoot{ The errors listed here, for this work and for OR08, are the standard  errors  of the mean ($\sigma_{obs}/\sqrt{N^{\circ} stars}$).}
\end{table}
%----------------------------------------------------------------

%----------------------------------------------------------------
%----------------------------------------------------------------
\begin{figure*}
\centering
  \includegraphics[width=6.9in,height=3.8in]{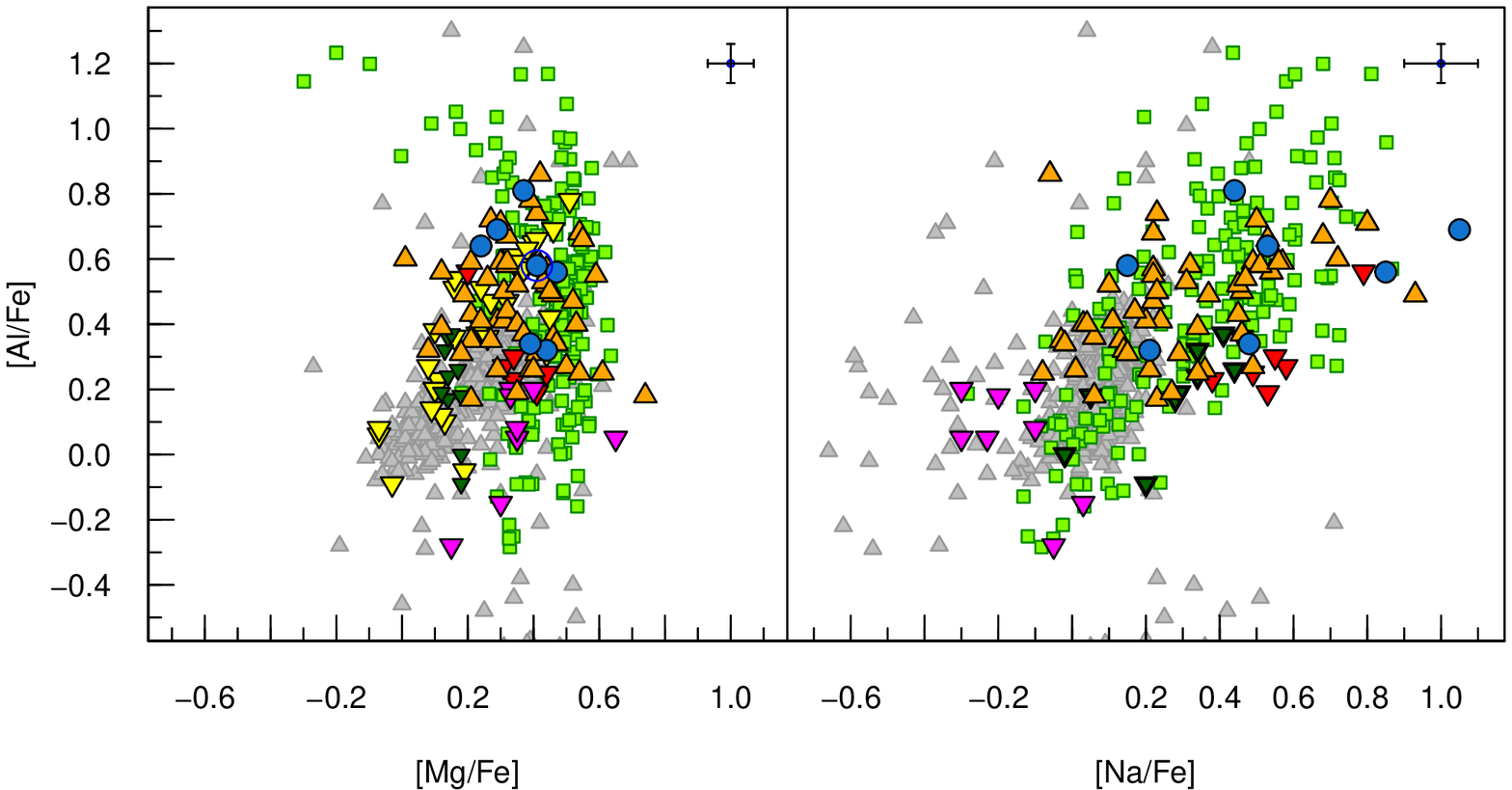}
  \caption{[Mg/Fe], [Na/Fe]  vs [Al/Fe]. Filled blue circles are our data for NGC 6440, filled red triangles :  NGC6441 \citep{gratton06},  Filled Yellow triangles: Terzan 5 \citep{origlia11}, filled dark green triangles: NGC 6553 \citep{tang17}, filled magenta triangles: HP1 \citep{barbuy16}, filled orange triangles: bulge field stars \citep{lecureur07}, filled green square: GCs from \citep{carretta09b},  filled grey triangles : Halo and Disk fields stars \citep{fulbright00,reddy03,reddy06,barklem05,cayrel04}.}
  \label{mg-al-na}
 \end{figure*}
%----------------------------------------------------------------
%----------------------------------------------------------------

   \begin{figure}[htb]
%\centering
  \includegraphics[width=3.5in,height=3.7in]{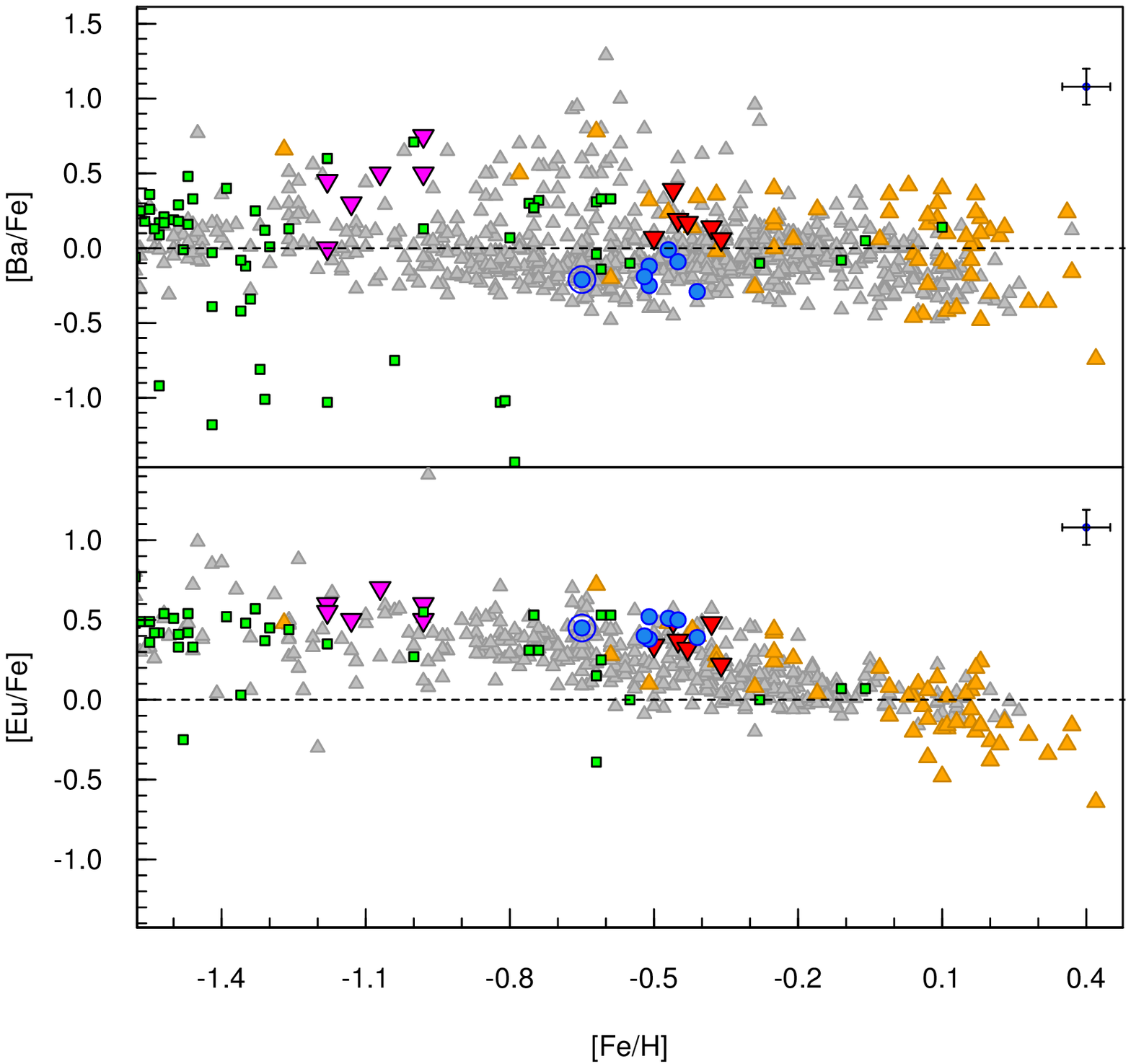}
  \caption{[Eu, Ba/Fe] vs [Fe/H]. Filled blue circles are our data for NGC 6440, filled red triangles: NGC 6441 \citep{gratton06},  filled magenta triangles: HP1 \citep{barbuy16}, filled orange triangles: bulge field stars \citep{vanderswaelmen16}, filled green square: GCs from  \citet{pritzl05}, filled  grey triangles: Halo and disk stars \citep{fulbright00,francois07,reddy06,barklem05,venn04}.}

  \label{heavy}
 \end{figure}
%----------------------------------------------------------------
%
 \begin{figure} [htb]
\centering
  \includegraphics[width=3.5in,height=3.7in]{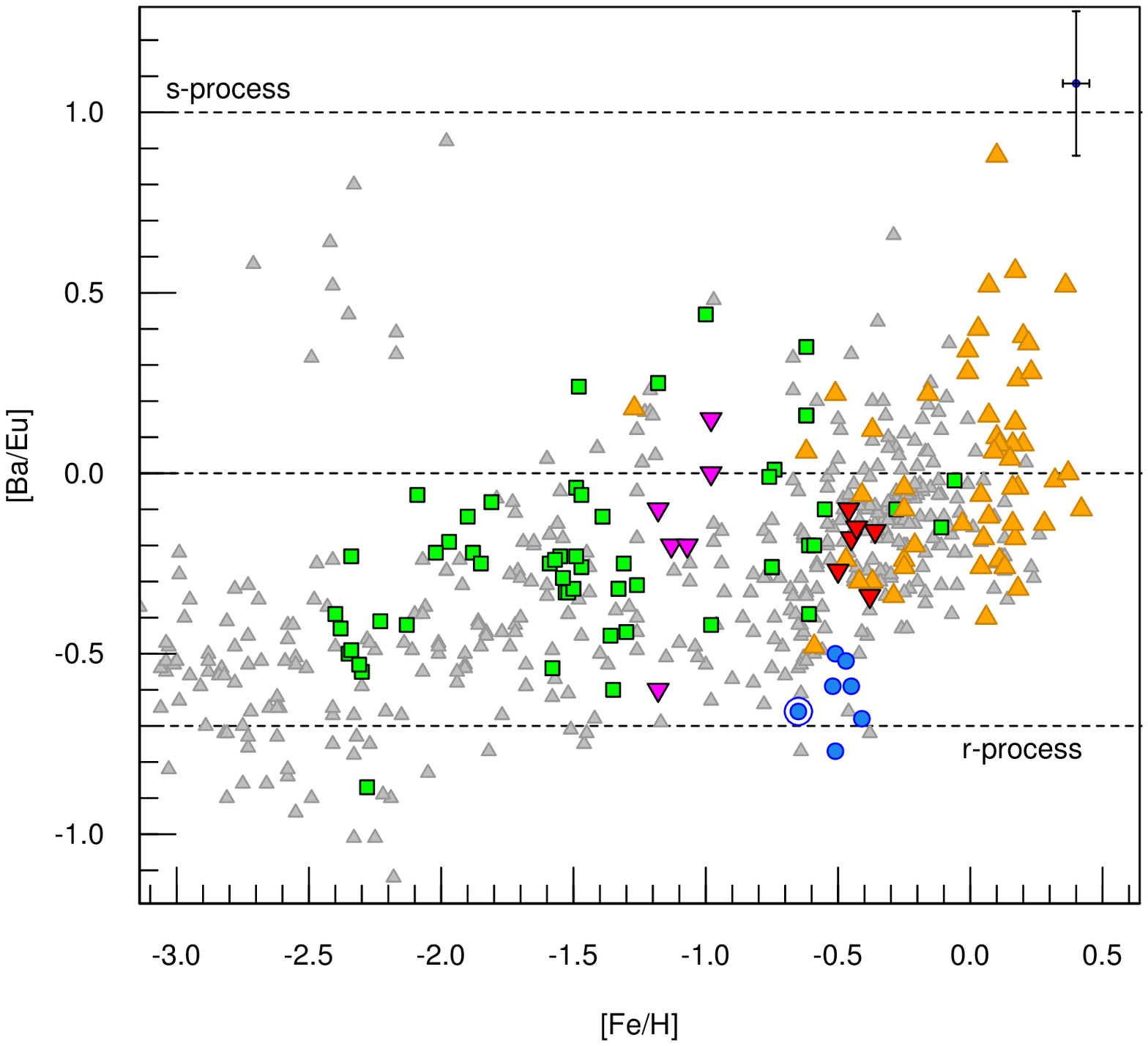}
  \caption{[Ba/Eu] vs [Fe/H]. Filled blue circles are our data for NGC 6440, filled red triangles: NGC 6441 \citep{gratton06},  filled magenta triangles: HP1 \citep{barbuy16}, filled orange triangles: bulge field stars \citep{vanderswaelmen16}, filled green square: GCs from  \citet{pritzl05}, filled  grey triangles: Halo and disk stars \citep{fulbright00,francois07,reddy06,barklem05,venn04}.}

  \label{baeu}
 \end{figure}
%

%----------------------------------------------------------------
%______________________________________________________________
\subsection{Mg-Al and Na-Al}

A Mg-Al anticorrelation has been found in several galactic GCs by different authors, particularly by \citet{carretta09a} in a large sample. Aluminium has shown a particularly high spread in almost all the galactic GCs, especially those with a  metallicity lower than {[}Fe/H{]}=-1.1 dex \citep{meszaros15}.

We have  found no  Mg-Al anticorrelation in NGC6440. Although  aluminium shows a significant spread 
 ($\sigma_{obs}=0.18$), magnesium does not ($\sigma_{obs}=0.08$). In this respect, NGC 6440 follows the trend of  bulge field stars and galactic GCs of similar metallicity.
The spread in abundance of Mg and Al found by OR08 ($\sigma_{Mg}=0.03$ and $\sigma_{Al}=0.05$) is small compared with ours. However it should be noted that the individual errors on the aluminum abundance  in the study of OR08 are in the order of 0.14 dex.

With respect to other bulge GCs, we noticed that Terzan 5 follows the trend of  bulge field stars, but not that of Galactic GCs. And NGC 6553 and HP1 do not agree with either bulge field stars or galactic GCs.

We plot in Fig. \ref{mg-al-na} Na vs Al, because these two light elements have the highest spread in this GC, allowing us to distinguish more clearly to which generation each star from our sample belongs. In this plot we can see a clear difference in the trends of  disk and halo  stars with respect to bulge stars, highlighting again their difference in chemical evolution history. In addition, we note a good agreement between  bulge stars and Galactic GCs including NGC6440. However, NGC 6553, NGC6441, and especially HP1, do not show such a good agreement with  bulge or galactic GCs.

%______________________________________________________________
\subsection{Heavy elements}

The heavy elements are produced by successive capture of neutrons through two processes: slow and rapid. In the first process (s-process), the neutron capture time is longer than the beta decay lifetime, for example during the AGB phase. On the other hand, the r-process occurs when the neutron capture time is much shorter that the beta-decay lifetime, for example during the SNeII. Therefore, the analysis of these elements undoubtedly helps to better understand the processes involved in the formation of the MP in the GCs, because they are very good tracers of stellar nucleosynthesis.

The production of s-elements is associated mainly with low-mass AGB stars (\citealt{gallino98}; \citealt{straniero06}), which can produce heavy elements like Ba. The r-process is associated mainly with SNe II explosions, with an important contribution of alpha elements and to a less extent iron-peak elements.\\

We measure two heavy elements in NGC 6440: Ba and Eu. Analysing the abundance of Ba for NGC 6440 we find  that the mean  is slightly sub-solar ({[}Ba/Fe{]}=-0.17 dex), although there is consistency with some bulge field stars with very low {[}Ba/Fe{]} (see Figure \ref{heavy}).The Bulge GCs NGC 6441 and HP1 show significant differences with  respect to NGC 6440, both GCs being\ super-solar in their content of Ba and following the trend of the bulge field stars. [Ba/Fe] in NGC 6440 is a bit lower than this  trend for  its  metallicity, indicating a low production of s-elements, which is associated with a  low  early pollution by AGB stars in NGC 6440.\\

In the case of {[}Eu/Fe{]}, there is a marked decrease with {[}Fe/H{]} for bulge stars. The decrease is due to to the production of iron by SN Ia after a time delay of 100 Myr to 1 Gyr \citep{ballero07}. The abundance of {[}Eu/Fe{]} in NGC 6440 shows values more in line with the trend of the bulge and in very good agreement with NGC 6441.

As  explained above, the trend of {[}Ba/Eu{]} vs {[}Fe/H{]} is very sensitive to the relative importance of s-process vs r-process. In the case of NGC 6440 we have found a
value of {[}Ba/Eu{]} which is very under-solar, and close to pure r-process (see Figure \ref{baeu}). Thus, in NGC 6440, heavy elements appear to have been mainly produced by explosive events like core collapse SNe.

%----------------------------------------------------------------
\section{Origin of NGC 6440}
NGC 6440 shows a good agreement with the bulge chemical pattern, indicating a common origin. Only some iron-peak elements (Co and Mn) are a bit higher  and [Ba/Eu] a bit lower  with respect to the Bulge field stars. This could indicate  that the place of formation of NGC 6440 suffered a more extended SN pre-enrichment compared with the bulk of the field stars and that it was born at the very beginning of the Bulge history when the contamination by AGB stars did not start yet. This may also help explain its relatively high alpha abundances compared to bulge objects of similar metallicity.

 NGC~6440 joins the other bulge GCs in the trend of alpha-element enrichment patterns for bulge stars, although with some dispersion. In particular, NGC~6440 seems to not present a spread in oxygen. Apparently bulge GCs have different ratios of first and second generation stars, maybe due to the harsh environment of the bulge to which  these clusters have been exposed  for a large period of their lives. More stars are needed to confirm this hypothesis.

There are no detailed studies on dynamics of NGC 6440. Proper motions are needed to complement our radial velocities and derive the orbit of this cluster to check whether it is actually confined to the bulge as the chemical analysis seems to indicate.

%______________________________________________________________
\section{Conclusions}
In this paper we have presented a detailed chemical analysis of NGC 6440 for seven of its RGB stars  with an S/N between 25-30 at 650\,nm. We measured  abundances of 14 chemical elements using high resolution spectra and we performed an accurate error analysis. This has allowed us to make a detailed comparison  with several important components of  the Milky Way (bulge field stars, disk field stars, halo field stars, galactic GCs), from a chemical point of view. The main findings are:
\begin{itemize}
         \item We find a mean  iron  abundance of [Fe/H]=-0.50$\pm$0.03 dex, in agreement with OR08. Like OR08, we did not find a significant spread in iron, although our sample is small;
       \item  Oxygen measurements for NGC 6440 show a mean [O/Fe]= 0.52$\pm$0.10 dex , one of the highest among Galactic GCs. The other alpha elments are also generally quite enriched;
      \item    Although Na shows an intrinsic spread, there is not a clear Na-O anticorrelation, since Oxygen does not show a significant spread, in   contrast with most  Galactic GCs; 
        \item  We  have found no Mg-Al anticorrelation but did detect an intrinsic spread in Al;    
     \item We found MP  in this cluster associated with the spread in   Na and Al;
  
    \item  $[$Ba/Eu$]$ suggests a low contribution by AGB stars to the gas from which  NGC 6440  formed;
        \item  Analysis of alpha, iron-peak and heavy elements indicates a strong and early contamination by SNe II;
        \item In general, the chemical abundances of stars in  NGC 6440 show good agreement with those of  bulge fields star in most of the elements analyzed in this paper, indicating a common formation and evolution process.
         \end{itemize}

  It is interesting to note that the comparison performed with other Bulge GCs such as NGC6441, HP1, NGC6553, Terzan 5 and  Terzan 4 shows both concordances and  discrepancies. For $\alpha$-elements we notice that  all  bulge GCs follow very nicely the trend with metallicity  of bulge field stars. Conversely, the  O-Na, Mg-Al and Na-Al plots, where we should see the typical inhomogeneities that characterize other galactic GCs, show  only a spread in Na but not a clear O-Na anticorrelation like in the case of HP1. This could suggest that bulge GCs underwent   different chemical evolution histories, but we need a larger sample of bulge GCs with detailed chemical measurements to reach a firm conclusion.

\begin{acknowledgements}
This work is Based on observations collected at the European Organisation for Astronomical Research in the Southern Hemisphere
under ESO programme ID 093.D-0286.
We gratefully acknowledge support from the Chilean BASAL   Centro de Excelencia en Astrof\'{i}sica
y Tecnolog\'{i}as Afines (CATA). C.M. is supported  by CONICYT (Chile)  through
 Programa Nacional de Becas de Doctorado  2014  (CONICYT-PCHA/Doctorado Nacional/2014-21141057). R.E.C. acknowledges support from Gemini-CONICYT for Project 32140007. We
would also like to thank the referee for his valuable comments and suggestions.
\end{acknowledgements}

%-------------------------------------------------------------------
\bibliographystyle{aa} % style aa.bst
\bibliography{biblio.bib}

\begin{thebibliography}{72}
\expandafter\ifx\csname natexlab\endcsname\relax\def\natexlab#1{#1}\fi

\bibitem[{{Ballero} {et~al.}(2007){Ballero}, {Matteucci}, {Origlia}, \&
  {Rich}}]{ballero07}
{Ballero}, S.~K., {Matteucci}, F., {Origlia}, L., \& {Rich}, R.~M. 2007, \aap,
  467, 123

\bibitem[{{Barbuy} {et~al.}(2016){Barbuy}, {Cantelli}, {Vemado}, {Ernandes},
  {Ortolani}, {Saviane}, {Bica}, {Minniti}, {Dias}, {Momany}, {Hill},
  {Zoccali}, \& {Siqueira-Mello}}]{barbuy16}
{Barbuy}, B., {Cantelli}, E., {Vemado}, A., {et~al.} 2016, \aap, 591, A53

\bibitem[{{Barbuy} {et~al.}(2013){Barbuy}, {Hill}, {Zoccali}, {Minniti},
  {Renzini}, {Ortolani}, {G{\'o}mez}, {Trevisan}, \& {Dutra}}]{barbuy13}
{Barbuy}, B., {Hill}, V., {Zoccali}, M., {et~al.} 2013, \aap, 559, A5

\bibitem[{{Barklem} {et~al.}(2005){Barklem}, {Christlieb}, {Beers}, {Hill},
  {Bessell}, {Holmberg}, {Marsteller}, {Rossi}, {Zickgraf}, \&
  {Reimers}}]{barklem05}
{Barklem}, P.~S., {Christlieb}, N., {Beers}, T.~C., {et~al.} 2005, \aap, 439,
  129

\bibitem[{{Carretta} {et~al.}(2009{\natexlab{a}}){Carretta}, {Bragaglia},
  {Gratton}, {D'Orazi}, \& {Lucatello}}]{carretta09c}
{Carretta}, E., {Bragaglia}, A., {Gratton}, R., {D'Orazi}, V., \& {Lucatello},
  S. 2009{\natexlab{a}}, \aap, 508, 695

\bibitem[{{Carretta} {et~al.}(2009{\natexlab{b}}){Carretta}, {Bragaglia},
  {Gratton}, \& {Lucatello}}]{carretta09a}
{Carretta}, E., {Bragaglia}, A., {Gratton}, R., \& {Lucatello}, S.
  2009{\natexlab{b}}, \aap, 505, 139

\bibitem[{{Carretta} {et~al.}(2010{\natexlab{a}}){Carretta}, {Bragaglia},
  {Gratton}, {Lucatello}, {Bellazzini}, {Catanzaro}, {Leone}, {Momany},
  {Piotto}, \& {D'Orazi}}]{carretta10a}
{Carretta}, E., {Bragaglia}, A., {Gratton}, R.~G., {et~al.} 2010{\natexlab{a}},
  \apjl, 714, L7

\bibitem[{{Carretta} {et~al.}(2009{\natexlab{c}}){Carretta}, {Bragaglia},
  {Gratton}, {Lucatello}, {Catanzaro}, {Leone}, {Bellazzini}, {Claudi},
  {D'Orazi}, {Momany}, {Ortolani}, {Pancino}, {Piotto}, {Recio-Blanco}, \&
  {Sabbi}}]{carretta09b}
{Carretta}, E., {Bragaglia}, A., {Gratton}, R.~G., {et~al.} 2009{\natexlab{c}},
  \aap, 505, 117

\bibitem[{{Carretta} {et~al.}(2010{\natexlab{b}}){Carretta}, {Bragaglia},
  {Gratton}, {Recio-Blanco}, {Lucatello}, {D'Orazi}, \&
  {Cassisi}}]{carretta10b}
{Carretta}, E., {Bragaglia}, A., {Gratton}, R.~G., {et~al.} 2010{\natexlab{b}},
  \aap, 516, A55

\bibitem[{{Cayrel} {et~al.}(2004){Cayrel}, {Depagne}, {Spite}, {Hill}, {Spite},
  {Fran{\c c}ois}, {Plez}, {Beers}, {Primas}, {Andersen}, {Barbuy},
  {Bonifacio}, {Molaro}, \& {Nordstr{\"o}m}}]{cayrel04}
{Cayrel}, R., {Depagne}, E., {Spite}, M., {et~al.} 2004, \aap, 416, 1117

\bibitem[{{Cescutti} \& {Matteucci}(2011)}]{cescutti11}
{Cescutti}, G. \& {Matteucci}, F. 2011, \aap, 525, A126

\bibitem[{{Cescutti} {et~al.}(2008){Cescutti}, {Matteucci}, {Lanfranchi}, \&
  {McWilliam}}]{cescutti08}
{Cescutti}, G., {Matteucci}, F., {Lanfranchi}, G.~A., \& {McWilliam}, A. 2008,
  \aap, 491, 401

\bibitem[{{Cohen} {et~al.}(2017){Cohen}, {Moni Bidin}, {Mauro}, {Bonatto}, \&
  {Geisler}}]{cohen17}
{Cohen}, R.~E., {Moni Bidin}, C., {Mauro}, F., {Bonatto}, C., \& {Geisler}, D.
  2017, \mnras, 464, 1874

\bibitem[{{Cottrell} \& {Da Costa}(1981)}]{cottrell81}
{Cottrell}, P.~L. \& {Da Costa}, G.~S. 1981, \apjl, 245, L79

\bibitem[{{Da Costa} {et~al.}(2009){Da Costa}, {Held}, {Saviane}, \&
  {Gullieuszik}}]{dacosta09}
{Da Costa}, G.~S., {Held}, E.~V., {Saviane}, I., \& {Gullieuszik}, M. 2009,
  \apj, 705, 1481

\bibitem[{{D'Antona} {et~al.}(2002){D'Antona}, {Caloi}, {Montalb{\'a}n},
  {Ventura}, \& {Gratton}}]{dantona02}
{D'Antona}, F., {Caloi}, V., {Montalb{\'a}n}, J., {Ventura}, P., \& {Gratton},
  R. 2002, \aap, 395, 69

\bibitem[{{D'Antona} {et~al.}(2016){D'Antona}, {Vesperini}, {D'Ercole},
  {Ventura}, {Milone}, {Marino}, \& {Tailo}}]{dantona16}
{D'Antona}, F., {Vesperini}, E., {D'Ercole}, A., {et~al.} 2016, \mnras, 458,
  2122

\bibitem[{{de Mink} {et~al.}(2010){de Mink}, {Pols}, {Langer}, \&
  {Izzard}}]{demink10}
{de Mink}, S.~E., {Pols}, O.~R., {Langer}, N., \& {Izzard}, R.~G. 2010, in IAU
  Symposium, Vol. 266, Star Clusters: Basic Galactic Building Blocks Throughout
  Time and Space, ed. R.~{de Grijs} \& J.~R.~D. {L{\'e}pine}, 169--174

\bibitem[{{Decressin} {et~al.}(2007){Decressin}, {Meynet}, {Charbonnel},
  {Prantzos}, \& {Ekstr{\"o}m}}]{decressin07}
{Decressin}, T., {Meynet}, G., {Charbonnel}, C., {Prantzos}, N., \&
  {Ekstr{\"o}m}, S. 2007, \aap, 464, 1029

\bibitem[{{Dias} {et~al.}(2016{\natexlab{a}}){Dias}, {Barbuy}, {Saviane},
  {Held}, {Da Costa}, {Ortolani}, {Gullieuszik}, \& {V{\'a}squez}}]{dias16a}
{Dias}, B., {Barbuy}, B., {Saviane}, I., {et~al.} 2016{\natexlab{a}}, \aap,
  590, A9

\bibitem[{{Dias} {et~al.}(2015){Dias}, {Barbuy}, {Saviane}, {Held}, {Da Costa},
  {Ortolani}, {Vasquez}, {Gullieuszik}, \& {Katz}}]{dias15}
{Dias}, B., {Barbuy}, B., {Saviane}, I., {et~al.} 2015, \aap, 573, A13

\bibitem[{{Dias} {et~al.}(2016{\natexlab{b}}){Dias}, {Saviane}, {Barbuy},
  {Held}, {Da Costa}, {Ortolani}, \& {Gullieuszik}}]{dias16b}
{Dias}, B., {Saviane}, I., {Barbuy}, B., {et~al.} 2016{\natexlab{b}}, The
  Messenger, 165, 19

\bibitem[{{Dotter} {et~al.}(2008){Dotter}, {Chaboyer}, {Jevremovi{\'c}},
  {Kostov}, {Baron}, \& {Ferguson}}]{dotter08}
{Dotter}, A., {Chaboyer}, B., {Jevremovi{\'c}}, D., {et~al.} 2008, \apjs, 178,
  89

\bibitem[{{Ferraro} {et~al.}(2009){Ferraro}, {Dalessandro}, {Mucciarelli},
  {Beccari}, {Rich}, {Origlia}, {Lanzoni}, {Rood}, {Valenti}, {Bellazzini},
  {Ransom}, \& {Cocozza}}]{ferraro09}
{Ferraro}, F.~R., {Dalessandro}, E., {Mucciarelli}, A., {et~al.} 2009, \nat,
  462, 483

\bibitem[{{Ferraro} {et~al.}(2016){Ferraro}, {Massari}, {Dalessandro},
  {Lanzoni}, {Origlia}, {Rich}, \& {Mucciarelli}}]{ferraro16}
{Ferraro}, F.~R., {Massari}, D., {Dalessandro}, E., {et~al.} 2016, \apj, 828,
  75

\bibitem[{{Fran{\c c}ois} {et~al.}(2007){Fran{\c c}ois}, {Depagne}, {Hill},
  {Spite}, {Spite}, {Plez}, {Beers}, {Andersen}, {James}, {Barbuy}, {Cayrel},
  {Bonifacio}, {Molaro}, {Nordstr{\"o}m}, \& {Primas}}]{francois07}
{Fran{\c c}ois}, P., {Depagne}, E., {Hill}, V., {et~al.} 2007, \aap, 476, 935

\bibitem[{{Fulbright}(2000)}]{fulbright00}
{Fulbright}, J.~P. 2000, \aj, 120, 1841

\bibitem[{{Gallino} {et~al.}(1998){Gallino}, {Arlandini}, {Busso}, {Lugaro},
  {Travaglio}, {Straniero}, {Chieffi}, \& {Limongi}}]{gallino98}
{Gallino}, R., {Arlandini}, C., {Busso}, M., {et~al.} 1998, \apj, 497, 388

\bibitem[{{Gnedin} \& {Ostriker}(1997)}]{gnedin1997}
{Gnedin}, O.~Y. \& {Ostriker}, J.~P. 1997, \apj, 474, 223

\bibitem[{{Gnedin} {et~al.}(2002){Gnedin}, {Zhao}, {Pringle}, {Fall}, {Livio},
  \& {Meylan}}]{gnedin02}
{Gnedin}, O.~Y., {Zhao}, H., {Pringle}, J.~E., {et~al.} 2002, \apjl, 568, L23

\bibitem[{{Gonzalez} {et~al.}(2012){Gonzalez}, {Rejkuba}, {Zoccali}, {Valenti},
  {Minniti}, {Schultheis}, {Tobar}, \& {Chen}}]{gonzalez12}
{Gonzalez}, O.~A., {Rejkuba}, M., {Zoccali}, M., {et~al.} 2012, \aap, 543, A13

\bibitem[{{Gratton} {et~al.}(2004){Gratton}, {Sneden}, \&
  {Carretta}}]{gratton04}
{Gratton}, R., {Sneden}, C., \& {Carretta}, E. 2004, \araa, 42, 385

\bibitem[{{Gratton} {et~al.}(2012){Gratton}, {Carretta}, \&
  {Bragaglia}}]{gratton12}
{Gratton}, R.~G., {Carretta}, E., \& {Bragaglia}, A. 2012, \aapr, 20, 50

\bibitem[{{Gratton} {et~al.}(2007){Gratton}, {Lucatello}, {Bragaglia},
  {Carretta}, {Cassisi}, {Momany}, {Pancino}, {Valenti}, {Caloi}, {Claudi},
  {D'Antona}, {Desidera}, {Fran{\c c}ois}, {James}, {Moehler}, {Ortolani},
  {Pasquini}, {Piotto}, \& {Recio-Blanco}}]{gratton07}
{Gratton}, R.~G., {Lucatello}, S., {Bragaglia}, A., {et~al.} 2007, \aap, 464,
  953

\bibitem[{{Gratton} {et~al.}(2006){Gratton}, {Lucatello}, {Bragaglia},
  {Carretta}, {Momany}, {Pancino}, \& {Valenti}}]{gratton06}
{Gratton}, R.~G., {Lucatello}, S., {Bragaglia}, A., {et~al.} 2006, \aap, 455,
  271

\bibitem[{{Harris}(1996)}]{harris96}
{Harris}, W.~E. 1996, \aj, 112, 1487

\bibitem[{{Izzard} {et~al.}(2013){Izzard}, {de Mink}, {Pols}, {Langer}, {Sana},
  \& {de Koter}}]{izzard13}
{Izzard}, R.~G., {de Mink}, S.~E., {Pols}, O.~R., {et~al.} 2013, \memsai, 84,
  171

\bibitem[{{Johnson} {et~al.}(2008){Johnson}, {Pilachowski}, {Simmerer}, \&
  {Schwenk}}]{johnson08}
{Johnson}, C.~I., {Pilachowski}, C.~A., {Simmerer}, J., \& {Schwenk}, D. 2008,
  \apj, 681, 1505

\bibitem[{{Johnson} {et~al.}(2014){Johnson}, {Rich}, {Kobayashi}, {Kunder}, \&
  {Koch}}]{johnson14}
{Johnson}, C.~I., {Rich}, R.~M., {Kobayashi}, C., {Kunder}, A., \& {Koch}, A.
  2014, \aj, 148, 67

\bibitem[{{Krause} {et~al.}(2013){Krause}, {Charbonnel}, {Decressin}, {Meynet},
  \& {Prantzos}}]{krause13}
{Krause}, M., {Charbonnel}, C., {Decressin}, T., {Meynet}, G., \& {Prantzos},
  N. 2013, \aap, 552, A121

\bibitem[{{Kurucz}(1970)}]{kurucz70}
{Kurucz}, R.~L. 1970, SAO Special Report, 309

\bibitem[{{Lecureur} {et~al.}(2007){Lecureur}, {Hill}, {Zoccali}, {Barbuy},
  {G{\'o}mez}, {Minniti}, {Ortolani}, \& {Renzini}}]{lecureur07}
{Lecureur}, A., {Hill}, V., {Zoccali}, M., {et~al.} 2007, \aap, 465, 799

\bibitem[{{Marcolini} {et~al.}(2009){Marcolini}, {Gibson}, {Karakas}, \&
  {S{\'a}nchez-Bl{\'a}zquez}}]{marcolini09}
{Marcolini}, A., {Gibson}, B.~K., {Karakas}, A.~I., \&
  {S{\'a}nchez-Bl{\'a}zquez}, P. 2009, \mnras, 395, 719

\bibitem[{{Marino} {et~al.}(2011{\natexlab{a}}){Marino}, {Milone}, {Piotto},
  {Villanova}, {Gratton}, {D'Antona}, {Anderson}, {Bedin}, {Bellini},
  {Cassisi}, {Geisler}, {Renzini}, \& {Zoccali}}]{marino11a}
{Marino}, A.~F., {Milone}, A.~P., {Piotto}, G., {et~al.} 2011{\natexlab{a}},
  \apj, 731, 64

\bibitem[{{Marino} {et~al.}(2011{\natexlab{b}}){Marino}, {Sneden}, {Kraft},
  {Wallerstein}, {Norris}, {da Costa}, {Milone}, {Ivans}, {Gonzalez},
  {Fulbright}, {Hilker}, {Piotto}, {Zoccali}, \& {Stetson}}]{marino11b}
{Marino}, A.~F., {Sneden}, C., {Kraft}, R.~P., {et~al.} 2011{\natexlab{b}},
  \aap, 532, A8

\bibitem[{{Marino} {et~al.}(2008){Marino}, {Villanova}, {Piotto}, {Milone},
  {Momany}, {Bedin}, \& {Medling}}]{marino08}
{Marino}, A.~F., {Villanova}, S., {Piotto}, G., {et~al.} 2008, \aap, 490, 625

\bibitem[{{Massari} {et~al.}(2014){Massari}, {Mucciarelli}, {Ferraro},
  {Origlia}, {Rich}, {Lanzoni}, {Dalessandro}, {Ibata}, {Lovisi}, {Bellazzini},
  \& {Reitzel}}]{massari14}
{Massari}, D., {Mucciarelli}, A., {Ferraro}, F.~R., {et~al.} 2014, \apj, 791,
  101

\bibitem[{{Mauro} {et~al.}(2012){Mauro}, {Moni Bidin}, {Cohen}, {Geisler},
  {Minniti}, {Catelan}, {Chen{\'e}}, \& {Villanova}}]{mauro12}
{Mauro}, F., {Moni Bidin}, C., {Cohen}, R., {et~al.} 2012, \apjl, 761, L29

\bibitem[{{Mauro} {et~al.}(2014){Mauro}, {Moni Bidin}, {Geisler}, {Saviane},
  {Da Costa}, {Gormaz-Matamala}, {Vasquez}, {Chen{\'e}}, {Cohen}, \&
  {Dias}}]{mauro14}
{Mauro}, F., {Moni Bidin}, C., {Geisler}, D., {et~al.} 2014, \aap, 563, A76

\bibitem[{{McWilliam} \& {Rich}(1994)}]{mcwilliam94}
{McWilliam}, A. \& {Rich}, R.~M. 1994, \apjs, 91, 749

\bibitem[{{M{\'e}sz{\'a}ros} {et~al.}(2015){M{\'e}sz{\'a}ros}, {Martell},
  {Shetrone}, {Lucatello}, {Troup}, {Bovy}, {Cunha},
  {Garc{\'{\i}}a-Hern{\'a}ndez}, {Overbeek}, {Allende Prieto}, {Beers},
  {Frinchaboy}, {Garc{\'{\i}}a P{\'e}rez}, {Hearty}, {Holtzman}, {Majewski},
  {Nidever}, {Schiavon}, {Schneider}, {Sobeck}, {Smith}, {Zamora}, \&
  {Zasowski}}]{meszaros15}
{M{\'e}sz{\'a}ros}, S., {Martell}, S.~L., {Shetrone}, M., {et~al.} 2015, \aj,
  149, 153

\bibitem[{{Minniti} {et~al.}(2010){Minniti}, {Lucas}, {Emerson}, {Saito},
  {Hempel}, {Pietrukowicz}, {Ahumada}, {Alonso}, {Alonso-Garcia}, {Arias},
  {Bandyopadhyay}, {Barb{\'a}}, {Barbuy}, {Bedin}, {Bica}, {Borissova},
  {Bronfman}, {Carraro}, {Catelan}, {Clari{\'a}}, {Cross}, {de Grijs},
  {D{\'e}k{\'a}ny}, {Drew}, {Fari{\~n}a}, {Feinstein}, {Fern{\'a}ndez
  Laj{\'u}s}, {Gamen}, {Geisler}, {Gieren}, {Goldman}, {Gonzalez}, {Gunthardt},
  {Gurovich}, {Hambly}, {Irwin}, {Ivanov}, {Jord{\'a}n}, {Kerins}, {Kinemuchi},
  {Kurtev}, {L{\'o}pez-Corredoira}, {Maccarone}, {Masetti}, {Merlo},
  {Messineo}, {Mirabel}, {Monaco}, {Morelli}, {Padilla}, {Palma}, {Parisi},
  {Pignata}, {Rejkuba}, {Roman-Lopes}, {Sale}, {Schreiber}, {Schr{\"o}der},
  {Smith}, {}, {Soto}, {Tamura}, {Tappert}, {Thompson}, {Toledo}, {Zoccali}, \&
  {Pietrzynski}}]{minniti10}
{Minniti}, D., {Lucas}, P.~W., {Emerson}, J.~P., {et~al.} 2010, \na, 15, 433

\bibitem[{{Origlia} {et~al.}(2013){Origlia}, {Massari}, {Rich}, {Mucciarelli},
  {Ferraro}, {Dalessandro}, \& {Lanzoni}}]{origlia13}
{Origlia}, L., {Massari}, D., {Rich}, R.~M., {et~al.} 2013, \apjl, 779, L5

\bibitem[{{Origlia} \& {Rich}(2004)}]{origlia04}
{Origlia}, L. \& {Rich}, R.~M. 2004, \aj, 127, 3422

\bibitem[{{Origlia} {et~al.}(2011){Origlia}, {Rich}, {Ferraro}, {Lanzoni},
  {Bellazzini}, {Dalessandro}, {Mucciarelli}, {Valenti}, \&
  {Beccari}}]{origlia11}
{Origlia}, L., {Rich}, R.~M., {Ferraro}, F.~R., {et~al.} 2011, \apjl, 726, L20

\bibitem[{{Origlia} {et~al.}(2008){Origlia}, {Valenti}, \& {Rich}}]{origlia08}
{Origlia}, L., {Valenti}, E., \& {Rich}, R.~M. 2008, \mnras, 388, 1419

\bibitem[{{Pritzl} {et~al.}(2005){Pritzl}, {Venn}, \& {Irwin}}]{pritzl05}
{Pritzl}, B.~J., {Venn}, K.~A., \& {Irwin}, M. 2005, \aj, 130, 2140

\bibitem[{{Reddy} {et~al.}(2006){Reddy}, {Lambert}, \& {Allende
  Prieto}}]{reddy06}
{Reddy}, B.~E., {Lambert}, D.~L., \& {Allende Prieto}, C. 2006, \mnras, 367,
  1329

\bibitem[{{Reddy} {et~al.}(2003){Reddy}, {Tomkin}, {Lambert}, \& {Allende
  Prieto}}]{reddy03}
{Reddy}, B.~E., {Tomkin}, J., {Lambert}, D.~L., \& {Allende Prieto}, C. 2003,
  \mnras, 340, 304

\bibitem[{{Renzini} {et~al.}(2015){Renzini}, {D'Antona}, {Cassisi}, {King},
  {Milone}, {Ventura}, {Anderson}, {Bedin}, {Bellini}, {Brown}, {Piotto}, {van
  der Marel}, {Barbuy}, {Dalessandro}, {Hidalgo}, {Marino}, {Ortolani},
  {Salaris}, \& {Sarajedini}}]{renzini15}
{Renzini}, A., {D'Antona}, F., {Cassisi}, S., {et~al.} 2015, \mnras, 454, 4197

\bibitem[{{Saviane} {et~al.}(2012){Saviane}, {da Costa}, {Held}, {Sommariva},
  {Gullieuszik}, {Barbuy}, \& {Ortolani}}]{saviane12}
{Saviane}, I., {da Costa}, G.~S., {Held}, E.~V., {et~al.} 2012, \aap, 540, A27

\bibitem[{{Smith} \& {Lambert}(1987)}]{smith87}
{Smith}, V.~V. \& {Lambert}, D.~L. 1987, \mnras, 226, 563

\bibitem[{{Sneden}(1973)}]{sneden73}
{Sneden}, C. 1973, \apj, 184, 839

\bibitem[{{Straniero} {et~al.}(2006){Straniero}, {Gallino}, \&
  {Cristallo}}]{straniero06}
{Straniero}, O., {Gallino}, R., \& {Cristallo}, S. 2006, Nuclear Physics A,
  777, 311

\bibitem[{{Tang} {et~al.}(2017){Tang}, {Cohen}, {Geisler}, {Schiavon},
  {Majewski}, {Villanova}, {Carrera}, {Zamora}, {Garcia-Hernandez}, {Shetrone},
  {Frinchaboy}, {Meza}, {Fern{\'a}ndez-Trincado}, {Mu{\~n}oz}, {Lin}, {Lane},
  {Nitschelm}, {Pan}, {Bizyaev}, {Oravetz}, \& {Simmons}}]{tang17}
{Tang}, B., {Cohen}, R.~E., {Geisler}, D., {et~al.} 2017, \mnras, 465, 19

\bibitem[{{Valenti} {et~al.}(2004){Valenti}, {Ferraro}, \&
  {Origlia}}]{valenti04}
{Valenti}, E., {Ferraro}, F.~R., \& {Origlia}, L. 2004, \mnras, 351, 1204

\bibitem[{{Van der Swaelmen} {et~al.}(2016){Van der Swaelmen}, {Barbuy},
  {Hill}, {Zoccali}, {Minniti}, {Ortolani}, \& {G{\'o}mez}}]{vanderswaelmen16}
{Van der Swaelmen}, M., {Barbuy}, B., {Hill}, V., {et~al.} 2016, \aap, 586, A1

\bibitem[{{Venn} {et~al.}(2004){Venn}, {Irwin}, {Shetrone}, {Tout}, {Hill}, \&
  {Tolstoy}}]{venn04}
{Venn}, K.~A., {Irwin}, M., {Shetrone}, M.~D., {et~al.} 2004, \aj, 128, 1177

\bibitem[{{Ventura} \& {D'Antona}(2006)}]{dantona06}
{Ventura}, P. \& {D'Antona}, F. 2006, \aap, 457, 995

\bibitem[{{Villanova} \& {Geisler}(2011)}]{villanova11}
{Villanova}, S. \& {Geisler}, D. 2011, \aap, 535, A31

\bibitem[{{Villanova} {et~al.}(2013){Villanova}, {Geisler}, {Carraro}, {Moni
  Bidin}, \& {Mu{\~n}oz}}]{villanova13}
{Villanova}, S., {Geisler}, D., {Carraro}, G., {Moni Bidin}, C., \&
  {Mu{\~n}oz}, C. 2013, \apj, 778, 186

\bibitem[{{Zoccali} {et~al.}(2014){Zoccali}, {Gonzalez}, {Vasquez}, {Hill},
  {Rejkuba}, {Valenti}, {Renzini}, {Rojas-Arriagada}, {Martinez-Valpuesta},
  {Babusiaux}, {Brown}, {Minniti}, \& {McWilliam}}]{zoccali14}
{Zoccali}, M., {Gonzalez}, O.~A., {Vasquez}, S., {et~al.} 2014, \aap, 562, A66

\end{thebibliography}

\end{document}